\documentclass[aps,prb,twocolumn,amsmath,amssymb,fleqn,superscriptaddress]{revtex4-1}
\usepackage{graphicx}
\usepackage{bm}
\usepackage{subfigure}
\usepackage{booktabs}
\usepackage{color}

\begin{document}

\title{Implications of second harmonic generation for hidden order in Sr$_2$CuO$_2$Cl$_2$}

\author{A. de la Torre}
\affiliation{Department of Physics, California Institute of Technology, Pasadena, CA 91125, USA}
\affiliation{Institute for Quantum Information and Matter, California Institute of Technology, Pasadena, CA 91125, USA}
\affiliation{Department of Physics, Brown University, Providence, RI 02912, USA}
\author{S. Di Matteo}
\affiliation{Univ Rennes, CNRS, IPR (Institut de Physique de Rennes) - UMR 6251, F-35708 Rennes, France}
\author{D. Hsieh}
\affiliation{Department of Physics, California Institute of Technology, Pasadena, CA 91125, USA}
\affiliation{Institute for Quantum Information and Matter, California Institute of Technology, Pasadena, CA 91125, USA}
\author{M. R. Norman}
\affiliation{Materials Science Division, Argonne National Laboratory, Lemont, IL  60439, USA}

\date{\today}

\begin{abstract} 
Sr$_2$CuO$_2$Cl$_2$ (SCOC) is a model undoped cuprate with $I4/mmm$ crystallographic symmetry, and a simple magnetic space group $C_Amca$ with associated magnetic point group $mmm1'$.  However, recent second harmonic spectroscopy in the antiferromagnetic phase has challenged this picture, suggesting instead a magnetic point group $4/mm'm'$ that co-exists with the antiferromagnetism and breaks the two orthogonal mirror planes containing the tetragonal $c$-axis. Here, we analyze the symmetry of SCOC in light of the second harmonic results, and discuss possible ground states that are consistent with the data.
\end{abstract}

\maketitle

\section{Introduction}

Second harmonic generation (SHG) is a powerful technique for detecting symmetry breaking,\cite{fiebig} with its high sensitivity to small structural distortions \cite{tor} and novel electronic order parameters.\cite{zhao1,zhao2,harter} Although a rigorous interpretation is still in development,\cite{fiebig,pershan,muto,sa,dmn,dmn2} the precise information that can be gathered about symmetries provides important hints for other investigations. As an example, the symmetry lowering from $I4_1/acd$ to $I4_1/a$ in Sr$_2$IrO$_4$ was detected by SHG \cite{tor} and confirmed and quantified by neutron diffraction.\cite{ye} 
Very recent SHG data \cite{torre} provided evidence for an order parameter in Sr$_2$CuO$_2$Cl$_2$ of magnetic symmetry $4/mm'm'$ . This is not compatible with the simple antiferromagnetic $mmm1'$ magnetic point group (MPG) identified by neutron diffraction,\cite{vaknin} and would lower this MPG to $mm'm'$. The present paper is focused on the analysis of this symmetry reduction.

Sr$_2$CuO$_2$Cl$_2$ (SCOC) is an insulating layered cuprate characterized by tetragonal $I4/mmm$ symmetry down to the lowest measured temperatures.\cite{vaknin} This tetragonality with its flat CuO$_2$ planes is likely stabilized by the apical chlorines with a rather long apical bond, thus suppressing the typical octahedral tilts seen in related cuprates like La$_2$CuO$_4$ (LCO). SCOC displays an antiferromagnetic (AFM) transition around $T_N\sim 260$ K with spins oriented along the (110) direction. It is worth noticing that due to the conservation of spin orientation under the body-centered translation of $I4/mmm$, time-reversal symmetry is also preserved in the MPG. Although the MPG for LCO is the same, the latter is orthorhombic due to octahedral tilts: the planar oxygens are no longer midway between Cu ions, thereby generating a Dzyaloshinskii-Moriya (DM) interaction that favors spin canting. This does not occur for SCOC ($I4/mmm$), as confirmed by the absence of x-ray magnetic circular dichroism (XMCD) contrast.\cite{deluca} These measurements also significantly constrain the existence of any ferromagnetic dipole component along the tetragonal $c$-axis.

Recent SHG measurements \cite{torre} found evidence for a $4/mm'm'$ magnetic order parameter (OP) associated with the breaking of two mirror planes containing the $c$-axis. This phenomenology cannot be explained with our previous knowledge of SCOC. Such an order parameter not only is unrelated to the antiferromagnetic $mmm1'$ point group, it is also unrelated to any magnetostriction that might occur due to the onset of AFM order, as it would result in a symmetry reduction to an $mmm$ point group. The SHG data also ruled out a surface effect. This is consistent with He-scattering measurements off the (001)-surface of SCOC, which revealed no reduction of the crystallographic symmetry.\cite{helium} As such, the SHG data points to the presence of a novel bulk electronic OP, reducing the MPG from $mmm1'$ to $mm'm'$, the subgroup of $mmm1'$ consistent with the newly discovered OP of $4/mm'm'$ magnetic symmetry.\cite{torre} Along the same direction, thermal Hall measurements indicate the presence of chirality \cite{boulanger} that is not consistent with a $4/mmm1'$ MPG.

The physical realization of such an OP is however still unclear. On one side, intra-unit cell magnetic order has been identified in underdoped cuprates by neutron diffraction.\cite{fauque}  A magneto-chiral generalization of the so-called `loop current' order \cite{he1,aji,he2,scheurer} has the observed $4/mm'm'$ symmetry, as it is equivalent to an orbital ferromagnet along $c$, and may therefore represent the magnetic OP evidenced by the SHG experiment.\cite{torre}  Another possible origin is a higher-order parity-even magnetic multipole like a magnetic octupole,\cite{santini1,igarashi,note1} 
which, if they exhibited ferroic ordering, could be revealed by SHG. We remind that such magnetic multipoles are not revealed by XMCD and, for example in the case of NpO$_2$, do not lead to any structural change.\cite{santini1} It is the goal of the present paper to analyze in detail the theoretical framework of the SHG experiment, relate this to other data in the literature, and then outline microscopic models that are consistent with the data.

To this aim, we first focus on a detailed description of the crystal and magnetic structure of SCOC in Section II. 
In Section III, we analyze the geometry of the SHG experiment and fits to the data,\cite{torre} and discuss the nature of the SHG, in particular the $d-d$ excitations involved in the SHG process. In Section IV, we list possible symmetry breakings and determine the constraints on the excited states that are consistent with the SHG findings. In Section V, we discuss possible microscopic models, and offer some concluding thoughts.

\section{Magnetic symmetry groups}

\begin{figure}[ht]
\includegraphics[width=\columnwidth]{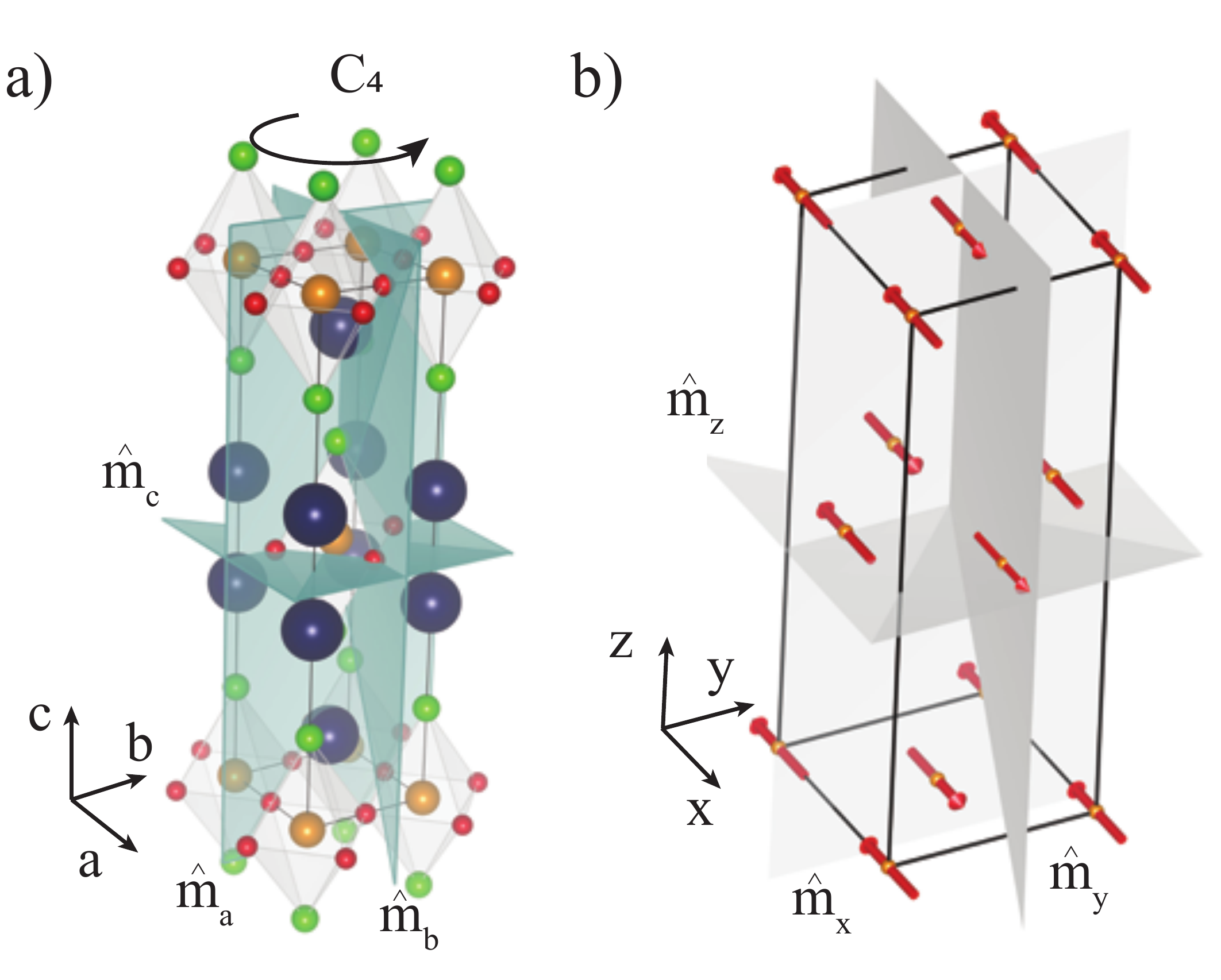}
\caption{a) The crystal structure of Sr$_2$CuO$_2$Cl$_2$, highlighting the symmetry elements. b) Magnetic structure, stressing the time reversal and mirror symmetries that might be broken as revealed by SHG.\cite{torre}  Cu is in gold, O in red, Cl in green, and Sr in blue.  The axes $x,y$ are rotated 45$^\circ$ relative to $a,b$.}
\label{fig1}
\end{figure}

Both x-ray \cite{grande} and neutron \cite{vaknin,miller} diffraction have shown that SCOC crystallizes in the body-centered tetragonal ($I4/mmm$) K$_2$NiF$_4$-type structure from 300 K down to 10 K.\cite{grande,miller,vaknin} The two Cu atoms per unit cell are in the 2a Wyckoff position,\cite{ITC} at $(0,0,0)$, characterized by the full point-group symmetry $4/mmm$, and related by the body-centered translation $(\frac{1}{2},\frac{1}{2},\frac{1}{2})$. The planar oxygens are at 4c sites, i.e., $(0,\frac{1}{2},0)$, and are characterized by the site symmetry $mmm$. The Cl and Sr ions are at 4e sites, i.e. $(0,0,z$), with $4mm$ symmetry. The absence of magnetism above $T_N$ makes time reversal $1'$ a symmetry of the material, so that the magnetic space group above $T_N$ is $I4/mmm1'$. This is shown in Fig.~\ref{fig1}a.

Below $T_N$, AFM order sets in as demonstrated by neutron diffraction.\cite{vaknin} The spin pattern is shown in Fig.~\ref{fig1}b, where the associated doubling of the unit cell is highlighted as well as the spin orientation along (110). The magnetic space group is orthorhombic $C_Amca$ in BNS notation \cite{Bilbao} ($F_Cmm'm'$ in OG notation \cite{Litvin}).  The corresponding MPG is $mmm1'$, so time reversal is still a symmetry of the material, with two of the three face-centering translations accompanied by a time reversal operator, and the other (the body-centered one of the original tetragonal cell) not.

Due to the intrinsic interest of the structural and magnetic point group in the present work, we specify below the symmetry elements, both above and below $T_N$, before considering the SHG experimental results. Above $T_N$, the 32 point group symmetry elements belonging to $4/mmm1'$ are the identity, the three mirror planes $\hat{m}_{a}$, $\hat{m}_{b}$, $\hat{m}_{c}$, the 4-fold axis along $c$, $\hat{4}_{c}^+$, $\hat{4}_{c}^-$, and the symmetry operations that can be derived from these, that is inversion ${\overline{1}}$, the three 2-fold axes $\hat{2}_{a}$, $\hat{2}_{b}$, $\hat{2}_{c}$, the two 4-fold roto-inversions $\overline{4}_{c}^+$, $\overline{4}_{c}^-$, the mirror planes perpendicular to the two diagonal directions in the plane, $\hat{m}_{x}$ and $\hat{m}_{y}$, and the two 2-fold axes along these directions, $\hat{2}_{x}$ and $\hat{2}_{y}$, where here $x$ and $y$ correspond to the orthorhombic directions in the magnetic phase. Of course, all the previous symmetry elements multiplied by the time reversal belong to $4/mmm1'$.

The nominal MPG associated with the AFM phase \cite{vaknin} is $mmm1'$. We remind that this MPG allows for two equivalent domains related by a $90^{\circ}$ rotation of the magnetic moment. Choosing the orthorhombic $x$ axis as the direction of the magnetic moment, as in Fig.~\ref{fig1}b, the following symmetry elements survive: the identity, $\hat{m}_{x}$, $\hat{m}_{y}$, $\hat{m}_{c}$, $\hat{2}_{x}$, $\hat{2}_{y}$, $\hat{2}_{c}$, ${\overline{1}}$, plus all of these symmetry elements multiplied by the time reversal $1'$. That is, the onset of the AFM order with moments pointing in the $x$-direction breaks all 4-fold symmetries, as well as mirror-planes and 2-fold axes oriented along the $a$ and $b$ tetragonal directions. Note, though, that there is another $C_Amca$ magnetic configuration that allows for staggered moments along $c$ instead.\cite{Bilbao}  This has not been observed in SCOC by neutron scattering.

In summary, SCOC has an $mmm1'$ MPG with antiferromagnetic moments along (110).  No evidence for canting has been seen,\cite{deluca} consistent with the fact that the paramagnetic $I4/mmm$ symmetry does not allow for a DM term.  But as mentioned above, SHG gives evidence for a co-existing magnetic order with $4/mm'm'$ symmetry.\cite{torre}  We turn to its description in Section IV.

\section{SHG experiment: energy level assignments }

Second-harmonic generation is a three-step process in the matter-radiation interaction, determined by two absorptions of a photon $\hbar\omega$ and the emission of a photon $2\hbar\omega$.\cite{fiebig} Its total scattering amplitude, $A_{SHG}$, can be written in quantum-mechanical terms using third-order perturbation theory as: 

\begin{align}
A_{SHG} = \sum_{l,n} \frac{\langle \Phi_g| H_I | \Phi_l \rangle \langle \Phi_l|  H_I| \Phi_n \rangle \langle \Phi_n|  H_I | \Phi_g \rangle}{(\Sigma_g-\Sigma_n)(\Sigma_g-\Sigma_l)}  
\label{ASHG}
\end{align}

\noindent where $\Phi_g$ is the initial state of the system (matter + radiation) and $\Phi_{n(l)}$ the intermediate states of the SHG process, of (matter + radiation) energies $\Sigma_g$ and $\Sigma_{n(l)}$, respectively.\cite{energy} $H_I$ is the matter-radiation interaction Hamiltonian, which is usually decomposed as a sum of electric dipole (E1), electric quadrupole (E2) and magnetic dipole (M1) contributions. The details of each term are given in Appendix A.1 and Section III of Ref.~\onlinecite{dmn}.
We remark here that the quantum-mechanical approach, though symmetry-wise equivalent to the semiclassical approach usually adopted in the optics literature,\cite{fiebig,boyd} allows for a deeper physical interpretation of the OP through the analysis of intermediate states.\cite{dmn2}

\begin{figure}[ht]
	\centering
	\includegraphics[width=\columnwidth]{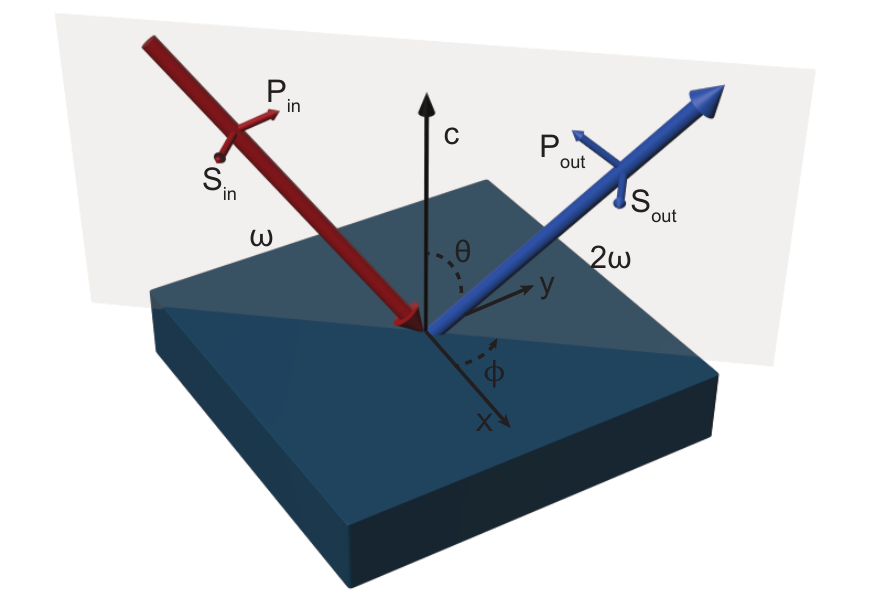}
	\caption{Geometry of the SHG experiment performed in Ref.~\onlinecite{torre}. The scattering plane is turned counterclockwise when viewed down the $c$ axis.}\label{fig2}
\end{figure}

We summarize the geometry of the experimental setup in Ref.~\onlinecite{torre} in the following. The experimental reference frame was such that the $x$-axis lies along the magnetic moment (45$^\circ$ from the $\vec{a}$ axis, as in Fig.~\ref{fig1}b), $z$ along the $\vec{c}$ axis and $y$ accordingly, as in Fig.~\ref{fig2}. We can then write the incoming/outgoing wave-vector $\vec{k}$ and electric field $\vec{\epsilon}$ (in $S$ and $P$ geometries) as:

\begin{subequations}
\begin{align}
\label{polax1}
& \vec{\epsilon}_S^{in} =  (\sin\phi,-\cos\phi,0)=\vec{\epsilon}_S^{out}  \\
\label{polax2}
& \vec{\epsilon}_P^{in} =  (\cos\theta \cos\phi,\cos\theta \sin\phi,\sin\theta) \\
\label{polax3}
& \vec{\epsilon}_P^{out} =  (-\cos\theta \cos\phi,-\cos\theta \sin\phi,\sin\theta)  \\
\label{polax4}
& \vec{k}^{in} =  (\sin\theta \cos\phi,\sin\theta \sin\phi,-\cos\theta)
\end{align}
\label{polax}
\end{subequations}

The angle $\phi$ is measured with respect to the direction of the magnetic moment and the angle $\theta$, the incidence angle, with respect to the normal to the CuO$_2$ plane. We remind that the $S$ geometry corresponds to the electric field perpendicular to the scattering plane and the $P$ geometry has the electric field within the scattering plane, as in Fig.~\ref{fig2}. In the experiment \cite{torre} the scattering plane was rotated counterclockwise around the $c$-axis. 
Eq.~\ref{polax1} shows that the electric field in $S$ geometry is independent of $\theta$ for both in and out channels. However, in the SS (S-in, S-out) geometry the experimental SHG signal is finite with $\theta \neq 0$, but zero for $\theta=0$.\cite{torre}
This necessarily excludes an E1-E1-E1 origin of the signal (such as from the surface), as the latter only depends on $\vec{\epsilon}_S^{in}$ and $\vec{\epsilon}_S^{out}$, which are independent of $\theta$. It also rules out any inversion breaking point group of SCOC, for which a $\theta$-independent E1-E1-E1 signal is expected in the SS geometry.
This implies that the detected signal in both the high-temperature (HT) and low-temperature (LT) phases must be either of E1-E1-M1 or E1-E1-E2 origin.

In order to determine the SCOC magnetic point group from the SHG data, it is important to discuss the symmetry of the intermediate transitions of the SHG process. In Ref~\onlinecite{torre}, the SHG signal collected at $\hbar\omega \simeq 1.5$ eV completely disappears when the photon energy is lowered to $1.0$ eV. This signifies resonant behavior. In fact, a non-resonant energy denominator in Eq.~\ref{ASHG} would have been of the kind $(E_{lg}+\hbar\omega)^{-1}$ or $(E_{ng}+2\hbar\omega)^{-1}$, with $E_{lg},E_{ng}>0$ the energy difference between the (matter only) excited and ground states. Therefore, it would have implied a small change in the signal when passing from $1.5$ to $1.0$ eV, but not its extinction. As such, we exclude non-resonant processes in the following discussion. 
Finally, we know from RIXS (resonant inelastic x-ray scattering) and optical data discussed below that $d-d$ excitations are present in the energy range 1.4 to 2.0 eV (below the charge transfer gap). Moreover, as discussed above, the SHG experiment cannot be explained in terms of an E1-E1-E1 process - so, an E2 or an M1 process must be present, i.e., a $d-d$ transition. We can therefore attribute the first resonant absorption at $\hbar\omega \simeq 1.5$ eV to a $d-d$ transition, through an E2 or M1 process.
The other two transitions are instead $d-p$ (E1) transitions.

\begin{figure*}[!ht]
\includegraphics[width=0.8\textwidth]{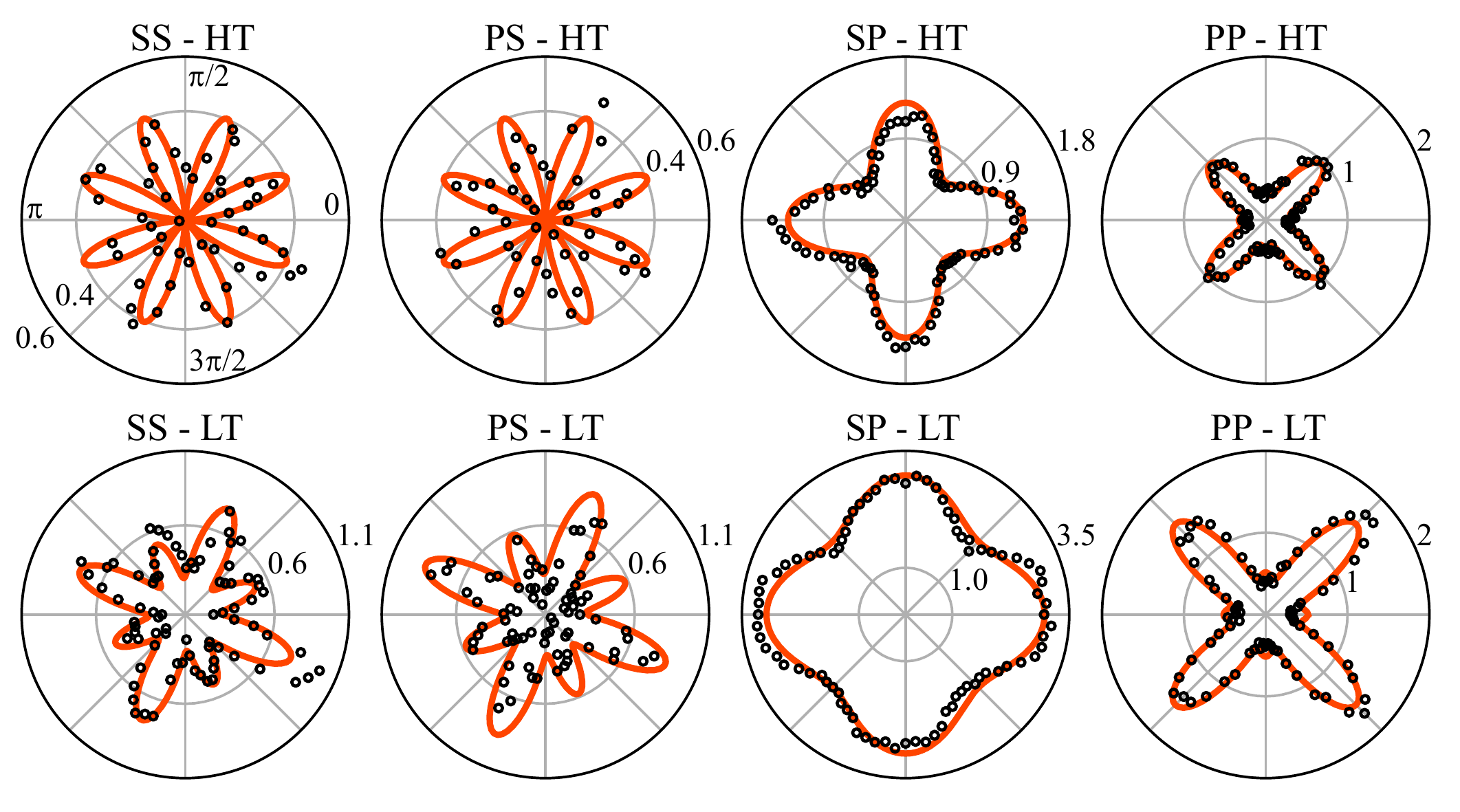}
\caption{Fits to the SHG azimuthal dependences from Ref.~\onlinecite{torre} for the four polarization geometries at 300 K (top row) and 20 K (bottom row).}
\label{fig3}
\end{figure*}

Unfortunately, the available literature does not allow us to unambiguously determine the symmetry of the $d$ orbitals for the intermediate state at 1.5 eV. 
A third harmonic generation (THG) experiment for SCOC \cite{schumacher} inferred an even parity 2$\omega$ transition in the energy range of 1.4 eV to 2.0 eV that was modeled as a transition from the $x^2-y^2$ ground state to tails of excited oscillators centered at 2 eV and above of $x^2-y^2$ symmetry, with another oscillator of $3z^2-r^2$ symmetry centered at 2 eV.
Whether other $d$ states are present or not is not known, since the geometry of the experiment was such that $xy$ and $xz,yz$ intermediate states could not be measured.  On the other hand, Raman data \cite{salamon} find a pronounced transition with $A_{2g}$ symmetry in Gd$_2$CuO$_4$ at 1.5 eV.  This is consistent with a transition from $x^2-y^2$ to $xy$.  Unfortunately, Raman data for SCOC have not been reported in this energy range, and a one-to-one assignment is not possible, as the energy of the $d-d$ excitations are sensitive to the planar Cu-O bond length, as well as the apical ions.  A power law relation of the $A_{2g}$ energy with in-plane Cu-O distance was found in Ref.~\onlinecite{salamon} for various cuprates, which, if literally applied to SCOC, would lead to an estimate of 1.35 eV for the $xy$ transition.\cite{calc}

Contrasting information about the $d-d$ transitions comes from RIXS experiments for SCOC. At the M$_3$ edge (3$p$ to 3$d$),\cite{kuiper} the $d-d$ transitions were estimated to be at 1.35 eV ($xy$), 1.5 eV ($3z^2-r^2$) and 1.7 eV ($xz,yz$), whereas at the L$_3$ edge (2$p$ to 3$d$) \cite{moretti} the estimates were instead 1.5 eV ($xy$), 1.84 eV ($xz,yz$), and 1.97 eV ($3z^2-r^2$).  The latter (based on an extensive data set) differ from previous L$_3$ estimates.\cite{ghiringhelli}  We note that the $d-d$ peaks are more pronounced in the $L_3$ data as compared to the $M_3$ data.  We also remark that the core hole, which differs for L$_3$ (2$p$) and M$_3$ (3$p$), could have a pronounced influence on the intermediate state energy levels, in contrast to optics where no such core hole perturbation is present.

In light of the above uncertainties, we consider all possibilities ($xy$, $xz/yz$ and $3z^2-r^2$) for the 1.5 eV transition in our analysis.

\section{Magnetic symmetry analysis of the SHG experiment}

With the expressions given in Appendix A for the M1 and E2 polarization dependences, it is possible to evaluate the SHG azimuthal dependence for any given MPG, by referring to the transformation properties of the polarization $P_{\alpha}$, the magnetization $M_{\alpha}$ and the electric quadrupole $Q_{\delta}$ transition operators discussed in Appendix B (see Table I).

We first summarize the details of the SHG azimuthal dependence reported in Ref~\onlinecite{torre} for $\theta = 11^{\circ}$, with the data and fits at two representative temperatures shown in Fig.~\ref{fig3}. The SHG HT intensity in the PS and SS channels can be fit by the expression $I_i(\phi) = |a_i\sin(4\phi)|^2$, with $i = PS, SS$. Interestingly, the experimental coefficients for the S-out channels are identical in magnitude within error bars: $a_{SS} = 0.64(1)$ and $a_{PS} = 0.64(1)$ at 300 K. Below $T_N$, the symmetry breaking observed in the azimuthal dependence of the S-out geometries signifies the onset of a new SHG channel. The LT functional form is $I_i(\phi) = |a_i\sin(4\phi) + b_i|^2$, with $a_i$ and $b_i$ in general being complex. In the SS geometry, the coefficients for the best fit to the data are $a_{SS} = 0.62(2)$, $b_{SS} = 0.50(1)$ with a relative phase angle between the two terms of $\gamma_{SS} = 102(2)^{\circ}$ at 20 K. Similar to the HT fit, the magnitude and relative phase of the PS coefficients matches that of the SS channel: $a_{SS} = 0.66(2)$, $b_{SS} = 0.53(2)$ and $\gamma_{SS} = 102(7)^{\circ}$ at 20 K.

In the P-out channels, no symmetry reduction is observed below $T_N$. In this case, the azimuthal dependence of the SHG is given by $I_i(\phi) = |a_i \cos(4\phi)+ b_i|^2$ with $i = SP, PP$, with the extracted fit parameters reflecting the linear increase of the SHG intensity with decreasing temperature in these channels. At 300 K, $a_{PP} = -0.49(2)$, $b_{PP} = 0.64(1)$, $\gamma_{PP} = 60(7)^{\circ}$, and $a_{SP} = 0.22(6)$, $b_{SP} = 0.95(1)$, $\gamma_{SP} \approx 33(15)^{\circ}$. At 20 K, $a_{PP} = -0.66(2)$, $b_{PP}= 0.77(1)$, and $a_{SP} = 0.17(1)$, $b_{SP} = 1.60(1)$, with no change in the relative phase with temperature (within error bars).

In the next subsections, we list and comment on the results of the azimuthal dependences for the MPG of the HT phase and some relevant MPGs for the LT phase.

\subsection{High-temperature magnetic point group $4/mmm1'$}

We start with the HT $4/mmm1'$ MPG. Given that the charge transfer gap is 2 eV, we limit the analysis to the case of a $2\omega$ (3 eV) transition of E1 origin ($d$ to $p$), as the E1 transition will always dominate if it is allowed.
\vspace{0.2cm}

{\it E1-E1-M1 channel -} No signal is present in SS and PS geometries. In PP and SP geometries, only one term is allowed, characterized by a constant azimuth (no $\phi$ dependence). Such a term behaves like $\sin(2\theta) (\chi_{aHT}+\chi_{bHT})$ in PP and $\sin(2\theta) \chi_{bHT}$ in SP. Here $\chi_{aHT}=\Im[P_xP_zM_y-P_yP_zM_x]$ and $\chi_{bHT}=\Im[P_z(P_xM_y-P_yM_x)]$. As in Appendix B, we use the notation: 
$\Im[P_{\alpha}P_{\beta}M_{\gamma}] \equiv \sum_{n,l} \langle g|P_\alpha|l\rangle\langle l|P_\beta|n\rangle\langle n|M_\gamma|g\rangle\Delta_{l,n} - c.c.$, where $g$ is the matter ground state, $l,n$ represent intermediate states, $\Delta_{l,n}$ is the resonant denominator of Eq.~\ref{ASHG} and $c.c.$ stands for the complex conjugate. We use below also the equivalent notation with $\Re[P_{\alpha}P_{\beta}M_{\gamma}]$, defined by the addition of the complex conjugate (instead of the difference).

We remark that $\chi_{aHT}=\chi_{bHT}=0$ if we impose the further constraint that the ground state has $d_{x^2-y^2}$ symmetry, that is there is one $x^2-y^2$ hole in the  $d$ shell of Cu (as demonstrated in the last paragraph of Appendix C). So, in this case, there is no signal in the E1-E1-M1 channel for any of the four geometries. 
\vspace{0.2cm}

{\it E1-E1-E2 channel -}  Of all the terms allowed by the MPG and reported in Appendix B, the two constraints of no outgoing E2 transition and $d_{x^2-y^2}$ symmetry for the ground state further reduce them to: 

\begin{subequations}
\begin{align}
\label{assE2HT}
& A_{SS} \propto \sin\theta \sin(4\phi) \chi_{HT} \\
\vspace{0.2cm}
\label{apsE2HT}
& A_{PS} \propto -\sin\theta\cos^2\theta \sin(4\phi) \chi_{HT} \\
\vspace{0.2cm}
\label{aspE2HT}
& A_{SP} \propto \sin\theta\cos\theta[\sin^2(2\phi) \chi_{HT} + \chi_1 + \chi_2] \\  
\vspace{0.2cm}
\label{appE2HT}
& A_{PP} \propto \sin\theta\cos^3\theta[\cos^2(2\phi) \chi_{HT} + \chi_2 +\tan^2\theta \chi_3]
\vspace{0.2cm}
\end{align}
\label{aE2HT}
\end{subequations}

In the above expressions, we introduced the correlation function $\chi_{HT}\equiv \Re[(P_x^2-P_y^2)Q_{x^2-y^2}]$.
We remark that this correlation function has the same symmetry as a non-magnetic hexadecapole. The other three correlation functions, less important in what follows, are defined through the three constants (in azimuth) $\chi_1=\Re[P_zP_xQ_{xz}+P_zP_yQ_{yz}]$, $\chi_2= -3\Re[(P_x^2+P_y^2)Q_{3z^2-r^2}] $ and $\chi_3=3\Re[P_z^2Q_{3z^2-r^2}]$. 

The azimuthal dependences in the four geometries correctly reproduce the experimental data:\cite{torre} they reproduce both the four-fold periodicity in the PP and SP channels and the eight-fold periodicity of $\sin^2(4\phi)$ in the SS and PS channels induced by the $\chi_{HT}$ term.
The presence of the single function $\chi_{HT}$ to explain both the SS and PS azimuthal dependences naturally explains the experimental data $A_{SS} \sim A_{PS}$.  In this context, the experimental $\theta$ is small ($11^\circ$) so $\cos(\theta) \sim 1$.

\begin{figure}[ht]
	\centering
	\includegraphics[width=\columnwidth]{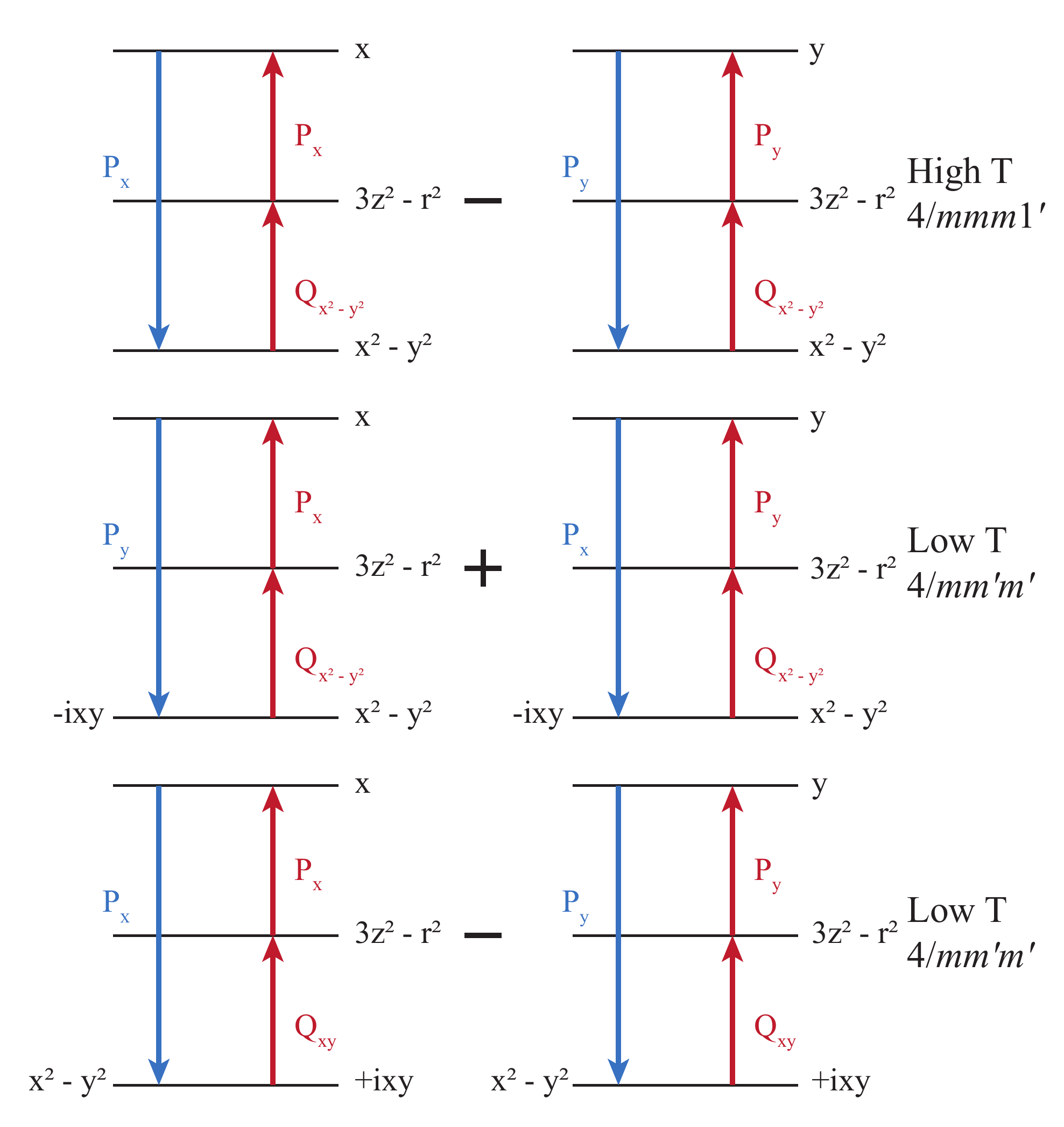}
	\caption{SHG process where the first transition is E2.  The top row refers to the HT phase, the bottom two rows to the LT phase.  Note the sign difference between $ixy$ in the second and third rows (i.e., $\chi_{LT3}= -\chi_{LT4}$), which plays an important role in the azimuthal dependence.}
	\label{fig4}
\end{figure}

We remark that, by using the selection rules detailed in Appendix C, the operator $Q_{x^2-y^2}$, applied to the $x^2-y^2$ ground state, can only lead to an intermediate state at 1.5 eV of $a_{1g}$ symmetry.  This implies that a $d-d$ transition to $3z^2-r^2$ is present at this energy.  This is illustrated in the top row of Fig.~\ref{fig4}.

\subsection{LT magnetic point groups}

Here we analyze the highest symmetry subgroups $4/mmm1'$ that share the symmetry of the experimental azimuthal scan: $4/mm'm'$ and $4/m1'$.
In fact, though the AFM arrangement observed by neutron scattering \cite{vaknin} breaks the four-fold symmetry of the total magnetic point group, this magnetic order is not directly observable by SHG (since the AFM order has finite momentum); SHG can only detect the magnetostriction induced by this order. Yet, magnetostriction (which would have orthorhombic symmetry) has never been reported for SCOC. So, as the LT experimental SHG signal keeps the HT four-fold symmetry and only breaks mirrors $\hat{m}_{a}$ and $\hat{m}_{b}$, it appears natural to look for an OP that keeps the highest symmetry sufficient to explain the data. Two cases are possible: if the OP has a magnetic origin (and therefore breaks the time-reversal symmetry $1'$), then the highest symmetry is $4/mm'm'$. If instead the OP has a structural origin, then time-reversal symmetry is not necessarily broken and the highest group compatible with the experimental SHG data is $4/m1'$. Apart from these MPGs, we also briefly analyze the antiferromagnetic MPG $mmm1'$ and its $mm'm'$ subgroup compatible with $4/mm'm'$, as well as $4'/mmm'$ (a proposed orbital current state in Ref.~\onlinecite{varma1}). 

The main aim of the following analysis is to explain the key features of the SHG experimental data:\cite{torre}
\vspace{0.2cm}

1) The LT contribution to PP and SP geometries must be identically zero, as there are no significant changes in PP and SP geometries when going from high to low temperatures. 
\vspace{0.2cm}

2) The LT contribution in the SS and PS channels is a constant, $\chi_{LT}$, that coherently sums with the HT E1-E1-E2 $\sin\theta\sin(4\phi)\chi_{HT}$ signal, so as to provide the observed azimuthal dependence $\propto f(\theta)|\chi_{HT}\sin(4\phi)+\chi_{LT}|^2$.
$\chi_{LT}$ can be complex relative to $\chi_{HT}$ (as seen by the fits presented earlier) and this can be appreciated from the general structure of Eq.~\ref{ASHG}. 
\vspace{0.2cm}

3) The $f(\theta)$ function contains at least one $\sin\theta$ factor, so as to explain that the experimental signal vanishes at normal incidence. 
\vspace{0.2cm}

We remark that, in all the cases discussed below, we further reduce the number of allowed SHG correlation functions found in Appendix B by imposing the constraint that the first transition must be M1 or E2 (i.e., a $d-d$ transition).

\subsubsection{LT OP of magnetic symmetry $4/mm'm'$}

{\it E1-E1-M1 channel -} With an incoming M1 transition, SS and PS geometries may have a non-zero signal, in both cases proportional to $\sin\theta$ and $\phi$ independent. It is associated with the correlation functions $\Re[(P_x^2+P_y^2)M_{z}]\equiv \chi_{LT1}$ for the SS case, and $\Re[P_xP_zM_{x}+P_yP_zM_{y}]\equiv \chi_{LT2}$ for the PS case. Moreover, for the $4/mm'm'$ MPG, the $d_{x^2-y^2}$ ground state is allowed to mix with $id_{xy}$. For this reason, $\chi_{aHT}$ and $\chi_{bHT}$ in SP and PP geometries are in general non-zero and can give a constant azimuth signal. We have: 

\begin{subequations}
\begin{align}
\label{assM1LT}
& A_{SS}\propto \sin\theta \chi_{LT1} \\
\vspace{0.2cm}
\label{apsM1LT}
& A_{PS}\propto -\sin\theta \chi_{LT2} \\
\vspace{0.2cm}
\label{aspM1LT}
& A_{SP}\propto \sin\theta\cos\theta \chi_{bHT} \\  
\vspace{0.2cm}
\label{appM1LT}
& A_{PP}\propto \sin\theta\cos\theta[\chi_{aHT} + \chi_{bHT}]
\vspace{0.2cm}
\end{align}
\label{aM1LT}
\end{subequations}

\begin{figure}[ht]
	\centering
	\includegraphics[width=\columnwidth]{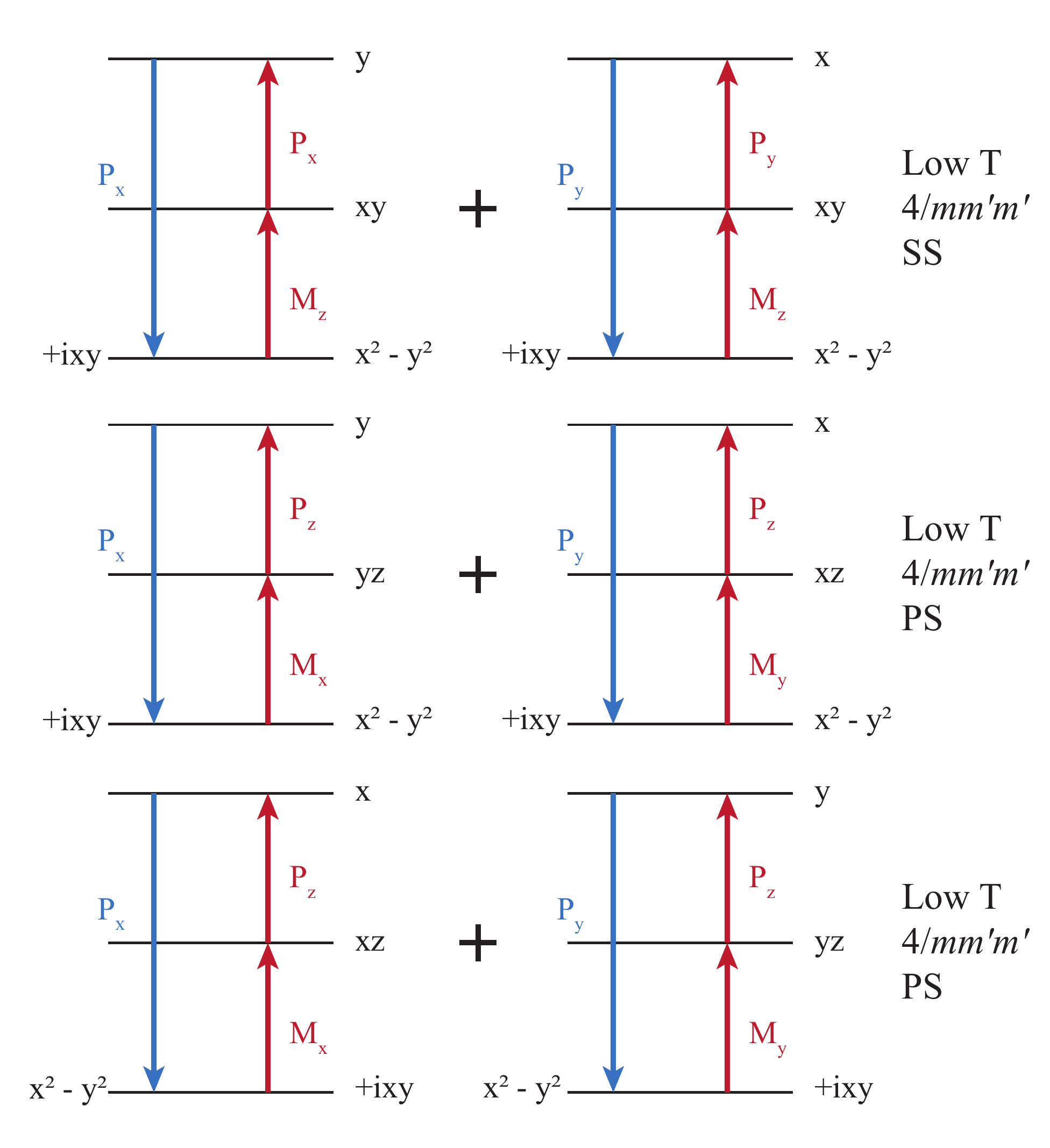}
	\caption{SHG process where the first transition is M1.  Note the difference in intermediate state between the top row ($xy$) and the bottom two rows ($xz,yz$), implying different strengths for these two processes.}
	\label{fig5}
\end{figure}

With these expressions, we can explain the above key features 1), 2) and 3) and fit the experimental data. Yet, we need two different correlation functions for the SS and PS geometries, $\chi_{LT1}$ and $\chi_{LT2}$. As SS and PS LT experimental data are quite similar in magnitude,\cite{torre} this would imply that these two correlation functions are coincidentally also similar. We remark that the correlation function $\chi_{LT2}$ is non-zero only if the intermediate state at 1.5 eV has $xz,yz$ symmetry, whereas the correlation function $\chi_{LT1}$ is non-zero only if the intermediate state at 1.5 eV has $xy$ symmetry, as illustrated in Fig.~\ref{fig5}.  As a consequence, one would expect these two processes to have different strengths at 1.5 eV, in contradiction to experiment.

\vspace{0.2cm}

{\it E1-E1-E2 channel -} Apart from the contribution of the HT parent group $4/mmm1'$, $\chi_{HT}$, the breaking of the time-reversal $1'$ symmetry leads to three more tensors, $\Im[(P_xP_y+P_yP_x)Q_{x^2-y^2}]\equiv \chi_{LT3}$, $\Im[(P_x^2-P_y^2)Q_{xy}]\equiv \chi_{LT4}$ and $\Im[P_xP_zQ_{yz}-P_yP_zQ_{xz}]\equiv 2\chi_{LT5}$, all time-reversal odd. By considering only the LT contributions, we have:
\vspace{0.2cm}

\begin{subequations}
\begin{align}
\label{assE2LT}
& A_{SS} \propto \sin\theta[-\sin^2(2\phi)\chi_{LT3} +\cos^2(2\phi) \chi_{LT4}]\\
\vspace{0.2cm}
\label{apsE2LT}
\begin{split}
& A_{PS} \propto \sin\theta\cos^2\theta\\
&[\sin^2(2\phi)\chi_{LT3} -\cos^2(2\phi) \chi_{LT4}- \chi_{LT5}] 
\end{split}
\\
\vspace{0.2cm}
\label{aspE2LT}
& A_{SP} \propto \frac{1}{2}\sin\theta\cos\theta\sin(4\phi)[\chi_{LT3} +\chi_{LT4}] \\
\vspace{0.2cm}
\label{appE2LT}
& A_{PP} \propto -\frac{1}{2}\sin\theta\cos\theta\sin(4\phi)[\chi_{LT3} +\chi_{LT4}] 
\vspace{0.2cm}
\end{align}
\label{aE2LT}
\end{subequations}

Though it appears as if this channel does not satisfy the required conditions (constant SS and PS and zero PP and SP LT signals), it turns out instead that it does, if the $d_{x^2-y^2}$ ground state is mixed with $id_{xy}$ (as expected for the $4/mm'm'$ MPG). In this case, as demonstrated in Appendix C, $\chi_{LT3}= -\chi_{LT4}$. This implies that the LT SHG signals in PP and SP geometries become zero and those in SS and PS geometries have the required constant azimuthal dependence. So, in this case also the E1-E1-E2 channel satisfies the key features 1), 2) and 3) and fits the experimental data. Moreover, in keeping with the similar magnitude of the LT SS and PS geometries, the agreement is reached in terms of only one parameter: $\chi_{LT}\equiv \chi_{LT3} = \Im[(P_x^2-P_y^2)Q_{xy}]$. We remark that, in this case, the action of the $Q_{xy}$ operator on the $ixy$ component of the ground state leads to the same intermediate state, of symmetry $3z^2-r^2$, as for the HT case. This, along with the relation of $\chi_{LT3}$ to $\chi_{LT4}$, is illustrated in Fig.~\ref{fig4}.  So, the E1-E1-E2 channel is a very plausible explanation of the experimental data. Here as well we need two hypotheses: the ground state of the form  $\alpha d_{x^2-y^2}+i\gamma d_{xy}$ (with $\gamma \ll \alpha$) and the absence of intermediate states of $xz,yz$ symmetry, that, as demonstrated in Appendix C, is needed to make $\chi_{LT5}=0$.  The latter is not strictly necessary - in principle it is sufficient to have $\chi_{LT5} \ll \chi_{LT}$.

We shall see below that the other MPGs will not explain the experimental data without further, implausible, hypotheses.

\subsubsection{LT OP of magnetic symmetry $4/m1'$}

{\it E1-E1-M1 channel -}  There are some analogies with the $4/mm'm'$ MPG: apart from the high-temperature term common to all subgroups of $4/mmm1'$, three correlation functions characterize this MPG: $\Im[P_xP_zM_{x}+P_yP_zM_{y}]$; $\Im[(P_x^2+P_y^2)M_z]$ and $\Im[P_z^2M_z]$. They are the imaginary counterparts (time-reversal even) of those analyzed above for the $4/mm'm'$ MPG. Their azimuthal scan is therefore the same: we find a constant contribution for SS and PS geometries, given by $\sin\theta \Im[(P_x^2+P_y^2)M_{z}]$ (SS case), and by $-\sin(\theta)\Im[P_xP_zM_{x}+P_yP_zM_{y}]$ (PS case). Also, we have the same drawback: the transitions take place through different intermediate states at 1.5 eV for the SS ($xy$) and PS ($xz,yz$) cases. 

{\it E1-E1-E2 channel -} The allowed terms of the $4/m1'$ MPG correspond to the real (time-reversal even) terms of those associated with $4/mm'm'$. So the azimuthal dependences are the same.
Yet, we remark on a fundamental difference between the two cases:  in this case ($4/m1'$), we need to impose $\Re[(P_xP_y+P_yP_x)Q_{x^2-y^2}] = - \Re[(P_x^2-P_y^2)Q_{xy}]$. As the symmetry of the $4/m1'$ MPG corresponds to a ground state of the form $d_{x^2-y^2}+\gamma d_{xy}$ with $\gamma$ real, we show in Appendix C that the latter condition cannot be met (rather, we have $\Re[(P_xP_y+P_yP_x)Q_{x^2-y^2}] = + \Re[(P_x^2-P_y^2)Q_{xy}]$). Therefore, we can discard this possibility.

\subsubsection{Other LT magnetic point groups: $4'/mmm'$, $mmm1'$ and $mm'm'$}

We included the MPG $4'/mmm'$ because it corresponds to the one that would be obtained if Varma's orbital current state $\Theta_I$ had been added to the SCOC HT phase. We exclude $\Theta_{II}$ from this discussion as it breaks inversion symmetry.\cite{varma1} 
The main difference between $4/mm'm'$ and $4'/mmm'$ in light of the SHG experiment is due to the fact that the two mirror planes of the $4'/mmm'$ MPG are not associated with time reversal, thereby failing to describe the experimental data, as only one mirror is broken and not both ($\hat{m}_{a}$ and $\hat{m}_{b}$). This is confirmed by the calculations reported in Appendix B showing that the calculated azimuthal dependences in the SS and PS geometries of the $4'/mmm'$ MPG are different from the experimental ones, so we can discard this MPG.

The nominal $mmm1'$ magnetic point group of the antiferromagnetic order does not allow to fulfill the constraints 1), 2) and 3) above, because the mirror symmetries $\hat{m}_x$ and $\hat{m}_y$ are not broken, contrary to the experimental data.  Detailed calculations \cite{torre} show that two-fold azimuthal dependences would appear, in keeping with the orthorhombic symmetry of this MPG.

The same would be true for the magnetic group $mm'm'$, that breaks the four-fold axis, whereas the experimental data show no measurable breaking of four-fold symmetry in all SS, PS, SP, and PP channels. This MPG would be the true magnetic point group of SCOC if we intersect the antiferromagnetic state with the $4/mm'm'$ magnetic OP revealed by SHG. 
Yet, having lost the four-fold symmetry, the $x$ and $y$ components belong to separate irreducible representations and for this reason, all terms containing the $x$ and $y$ components are characterized by a two-fold symmetry. An OP of this symmetry, therefore, cannot describe the four-fold symmetry of the SHG experiment.  
Presumably, the lack of observation of orthorhombicity (from magnetostriction) is either due to its weakness, or an equal population of both magnetic domains (i.e., either spins along $x$ or spins along $y$).
Moreover, the extra SHG signal in the LT phase also means that predominantly one domain of $4/mm'm'$ is present (as discussed below, orbital moments pointing along $\vec{c}$ as opposed to pointing along -$\vec{c}$).  Otherwise, the extra LT signal would either average out to zero, or its interference with the HT contribution would change sign depending on the domain, which was not observed in either spatial scans or thermal cycling of a given sample, or for different samples. The reason only a single domain of $4/mm'm'$  is seen remains an open question.

\section{Possible microscopic models and conclusions}  

From the discussion of the previous section, the most plausible magnetic symmetry of the OP detected by the SHG experiment \cite{torre} is $4/mm'm'$. The two scenarios outlined above were an SHG signal from (1) the E1-E1-E2 channel, with an intermediate state of $3z^2-r^2$ symmetry, or (2) the E1-E1-M1 channel, through intermediate states of $xy$ and $xz,yz$ symmetries.

Here we discuss some possible microscopic realizations of such an OP and suggest new experiments that might detect it. We first remind that SHG is only sensitive to \textit{ferro} OPs (that is, it is not sensitive to linear order to any finite $Q$ OPs). Therefore, we can neglect the symmetry reduction due to antiferromagnetic ordering in what follows. We know that the $4/mm'm'$ MPG breaks the time-reversal symmetry associated with the translation from a Cu site with a given spin (Cu$_1$) to another Cu site with opposite spin (Cu$_2$). 
We can see only three mechanisms that allow one to break the $1'$ symmetry from Cu$_1$ to Cu$_2$ while keeping the $4/mm'm'$ symmetry: 

\begin{figure}[ht]
\includegraphics[width=\columnwidth]{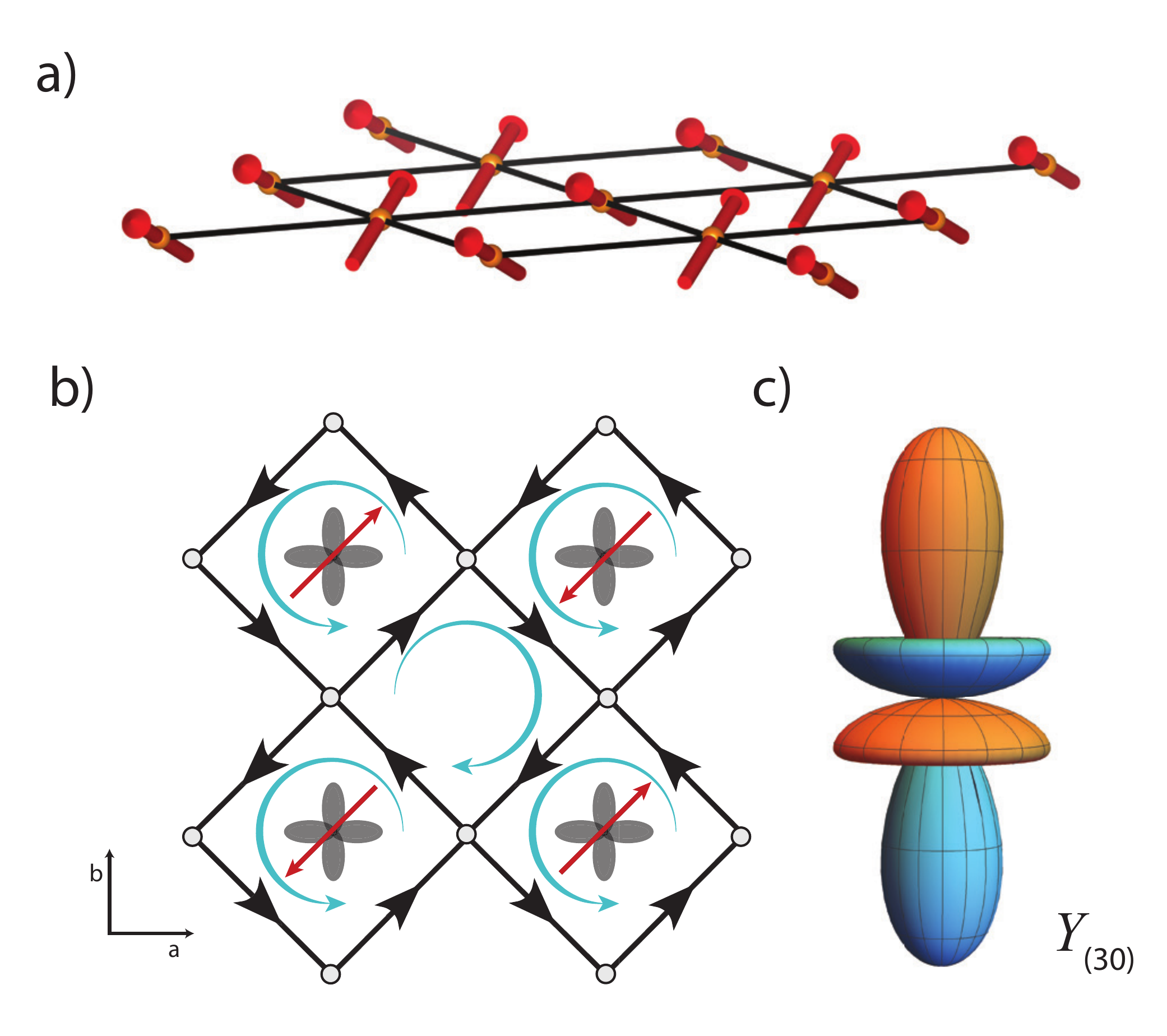}
\caption{ Illustration of the possible microscopic realizations of a $4/mm'm'$ order parameter in Sr$_2$CuO$_2$Cl$_2$: a) ferromagnetically ordered canted moments along the $c$-axis, b) magneto-chiral orbital currents around Cu sites and, c) ferroic ordering of a magnetic octupole $Y_{30}$.}
\label{fig6}
\end{figure}

1) The presence of a ferromagnetically ordered canted component of the magnetic moment along the $c$-axis (Fig.~\ref{fig6}a). Even though this possibility seems excluded by the XMCD measurements,\cite{deluca} there is either the possibility that SHG is more sensitive than XMCD or the possibility that canting is induced in the near surface region of the sample, that is, over a depth comparable to the photon absorption depth associated with 2$\omega$ ($\delta$ = 70 nm \cite{lovenich}).  We find this unlikely for two reasons. First, canting is not possible for an underlying $I4/mmm$ space group.  Second, when present (as in orthorhombic La$_2$CuO$_4$), the canting only becomes ferromagnetically aligned above a spin-flop field.

2) The presence of orbital currents around each Cu site (Fig.~\ref{fig6}b). Several models can be discussed in this framework, all characterized by circulating currents around Cu-sites that ferromagnetically order at both the Cu$_1$ and Cu$_2$ sites. In this way, they break the time-reversal symmetry while preserving the inversion symmetry. The simplest example involves currents flowing between the planar oxygen sites, leading to a magneto-chiral state.\cite{he1,aji,he2,scheurer}  In an effective one-band description, this mixes in an $id_{xy}$ component into the $d_{x^2-y^2}$ ground state as illustrated in Figs.~\ref{fig4} and \ref{fig5}.  The simplest way to see this is that these currents are equivalent to an orbital moment along $c$, and the $L_z$ operator leads to such mixing.  Then one has to presume that the resulting orbital moment was too small to have been observed by XMCD.

3) Higher-order magnetic multipoles, like a magnetic octupole (Fig.~\ref{fig6}c). A component of the magnetic octupole behaving like the zeroth component of a time-reversal odd spherical tensor of rank three ($Y_{30}$) would satisfy all the necessary symmetry constraints analyzed in the previous section if it exhibited ferroic ordering.  But why a magnetic octupole would arise in a material with spin-$\frac{1}{2}$ moments is not clear, even when invoking spin-orbit coupling.

Several experiments can be suggested to help resolve some of the questions raised by our work.
For instance, $4/mm'm'$ allows for an anomalous Hall effect \cite{Bilbao} as treated in Refs.~\onlinecite{he1,aji,he2,scheurer}. Although SCOC is an electrical insulator, high enough frequency measurements might allow for observation of the anomalous electrical Hall effect. Moreover, the thermal Hall analog is possible.  Although a large thermal Hall signal was reported in Ref.~\onlinecite{boulanger} which was attributed to a chiral contribution to the phonon thermal Hall effect (describable by a rank-3 tensor \cite{Bilbao}), to date, no anomalous signal has been reported. Besides the anomalous thermal Hall, there should also be a linear in field contribution to the longitudinal thermal conductivity. Polarized neutron scattering would be desirable to test for the presence of orbital currents or higher magnetic multipolar phases,\cite{fauque,santini1} which could also be detected in Sagnac-Faraday experiments.~\cite{Faraday}
A detailed angle of incidence dependence of the SHG response in the PS channel should be able to resolve the question of whether an E1-E1-M1 or an E1-E1-E2 process is responsible for the additional signal in the LT phase, as their $\theta$ dependences differ (Eqs.~4b and 5b). Finally, Raman experiments in the energy range of 1.4 to 2.0 eV would help determine what the symmetry of the $d-d$ excitations are.

In conclusion, we find that the additional second harmonic generation signal that appears in the magnetically ordered state of Sr$_2$CuO$_2$Cl$_2$ can be understood from an order parameter characterized by a $4/mm'm'$ magnetic point group that has the effect of mixing in an $id_{xy}$ component into the  $d_{x^2-y^2}$ ground state.  This OP behaves like an orbital ferromagnet.  We hope that future experiments can further elucidate this phenomenon.

\acknowledgments
Work at Caltech was supported by an ARO PECASE award W911NF-17-1-0204. 
Work at Argonne was supported by the Materials Sciences and Engineering Division, Basic Energy Sciences, Office of Science, US Department of Energy.

\appendix

\section{Expressions for the M1 and E2 polarization components}

First, it might be useful to remind how the decoupling of the E2 and M1 channels takes place for a generic fourth-rank tensor, like $\chi_{ijlm}$. Consider the specific HT case, with MPG $4/mmm1'$. In this case, the requirement that $\chi_{ijlm}$ is invariant under all symmetry operations of the group only allows those terms of $\chi_{ijlm}$ with an even number of $x$, $y$ or $z$ (the application of a two-fold axis on an odd number would change its sign). Moreover, the four-fold symmetry imposes that terms with equivalent $x$ and $y$ labels are equal (say, $\chi_{xyyx}=\chi_{yxxy}$). Limiting to the SS case (polarizations in the $xy$-plane), we are left with four terms: $\chi_{xxxx}=\chi_{yyyy}$, $\chi_{xxyy}=\chi_{yyxx}$, $\chi_{xyxy}=\chi_{yxyx}$, and $\chi_{xyyx}=\chi_{yxxy}$. Yet, this is true for any fourth-rank tensor, based only on its symmetry properties. The electric quadrupole SHG tensor is not {\it any} tensor, but a specific one - for example $\chi_{xyyx} = \sum_{n,l} \langle g|x|l\rangle\langle l|y|n\rangle\langle n|yx|g\rangle\Delta_{l,n}$ (here $g$ is the matter ground state, $l,n$ represent intermediate states, and $\Delta_{l,n}$ is the resonant denominator of Eq.~\ref{ASHG}). Therefore, the symmetry in the last two indexes provides a further equality, valid for the E2 channel, $\chi_{xyxy}=\chi_{xyyx}$. Analogously, it can be shown that, in the M1 channel, the tensor $\chi$ is antisymmetrized over the last two indexes: $\chi_{xyxy}=-\chi_{xyyx}$. The new tensor $\chi_{xyxy}-\chi_{xyyx}$ is often relabeled as $\chi_{xyz}$. In fact, as detailed in Refs.~\onlinecite{dmn,dmn2}, in the E1-E1-M1 channel, the $\chi$ tensor is scalarly coupled to the radiation terms $\vec{\epsilon}_{\alpha}\vec{\epsilon}_{\beta}(\vec{\epsilon}\times\vec{k})_{\gamma}$, where $\alpha$, $\beta$ and $\gamma$ represent any of $x$, $y$ and $z$. The vector product allows one to reduce the number of labels from four to three.  These considerations were first noted by Pershan.\cite{pershan}

In the particular geometrical configuration of Fig.~\ref{fig2}, the coupling to incoming and outgoing radiation for M1 transitions is expressed as: 

\begin{align}
& (\vec{\epsilon}\times\vec{k})_P^{in} =  (-\sin\phi,\cos\phi,0)=(\vec{\epsilon}\times\vec{k})_P^{out} \nonumber \\
& (\vec{\epsilon}\times\vec{k})_S^{in} =  (\cos\theta \cos\phi,\cos\theta \sin\phi,\sin\theta) \nonumber \\
& (\vec{\epsilon}\times\vec{k})_S^{out} =  (-\cos\theta \cos\phi,-\cos\theta \sin\phi,\sin\theta)
\label{polaxM1}
\end{align}

Coming back to the E1-E1-E2 channel, the tensor $\chi_{ijlm}$ is scalarly coupled to $\vec{\epsilon}_{\alpha}\vec{\epsilon}_{\beta}\{\vec{\epsilon},\vec{k}\}_{\delta}$ (see Refs.~\onlinecite{dmn,dmn2}), where $\{.,.\}_{\delta}$ means `symmetrization' and $\delta$ is shorthand notation for the five quadrupole terms, with the following order: $1 \rightarrow {x^2-y^2}$, $2 \rightarrow {3z^2-r^2}$, $3 \rightarrow {xy}$, $4 \rightarrow {xz}$ and $5 \rightarrow {yz}$. We remark that of the six symmetric components, the scalar term is zero, because it is coupled to $\vec{\epsilon}\cdot\vec{k}=0$.
In the particular geometrical configuration of Fig.~\ref{fig2}, the coupling to incoming and outgoing radiation for E2 transitions (with the above order) is expressed as:

\begin{align}
& \{\vec{\epsilon},\vec{k}\}_S^{in} =  [\sin\theta\sin(2\phi), 0,  -\sin\theta\cos(2\phi),  \nonumber \\
&   -\cos\theta\sin\phi, \cos\theta\cos\phi] \nonumber \\
& \{\vec{\epsilon},\vec{k}\}_S^{out} =  [\sin\theta\sin(2\phi), 0, -\sin\theta\cos(2\phi),  \nonumber \\
&  \cos\theta\sin\phi, -\cos\theta\cos\phi] \nonumber \\
& \{\vec{\epsilon},\vec{k}\}_P^{in} =  [\frac{1}{2}\sin(2\theta)\cos(2\phi), -\frac{3}{2}\sin(2\theta),  \nonumber \\
& \frac{1}{2}\sin(2\theta)\sin(2\phi), -\cos(2\theta)\cos\phi, -\cos(2\theta)\sin\phi] \nonumber \\
& \{\vec{\epsilon},\vec{k}\}_P^{out} =  [-\frac{1}{2}\sin(2\theta)\cos(2\phi), \frac{3}{2}\sin(2\theta),  \nonumber \\
& -\frac{1}{2}\sin(2\theta)\sin(2\phi), -\cos(2\theta)\cos\phi, -\cos(2\theta)\sin\phi]
\label{polaxE2}
\end{align}

\section{Calculation of the allowed tensors for each MPG}

Here we evaluate the allowed OPs for the HT $4/mmm1'$ MPG and some of its subgroups. We introduce the notation $P_{\alpha}$, $M_{\alpha}$, $Q_{\delta}$ for the transition operators of, respectively, the electric dipole, magnetic dipole and electric quadrupole, with $\alpha = x,y,z$ and $\delta=$ the five quadrupole components in the order listed in Appendix A. For example, $P_x=\langle g| x| l\rangle$ or $M_y=\langle n| L_y| g\rangle$ ($L_y$ is the $y$ component of the orbital angular momentum), or $Q_1=\langle n| x^2-y^2| g\rangle$. Even though these matrix elements clearly depend on the states ($| g\rangle$, $| n\rangle$, etc.), in the following we shall mainly be concerned with their geometrical transformation properties and, in order to lighten the notation, we shall not explicitly label the states, if not needed. However, when included in brackets, like $[P_xP_yM_z]$, it is shorthand notation for the whole SHG amplitude $\sum_{l,n}\Delta_{l,n}\langle g|P_x |l \rangle \langle l| P_y |n \rangle \langle n| L_{z}|g\rangle$, and therefore the order is important: $[P_xP_yM_z]\neq [P_yP_xM_z]$.

\begin{table}[ht!]
	\caption{Transformation table of the electric dipole, $\vec{P}$, the magnetic dipole, $\vec{M}$, and the electric quadrupole, Q$_{i}$, under the action of symmetry elements of the $4/mmm1'$ magnetic point group.}
	\centering
	\begin{ruledtabular}
		\begin{tabular}{c|ccc|ccc|ccccc|}
			 Sym. & P$_x$ & P$_y$ & P$_z$ & M$_x$ & M$_y$ & M$_z$ & Q$_1$ & Q$_2$ & Q$_3$ & Q$_4$ & Q$_5$ \\
			\colrule
	$\hat{1}$      & P$_x$ & P$_y$ & P$_z$ & M$_x$ & M$_y$ & M$_z$ & Q$_1$ & Q$_2$ & Q$_3$ & Q$_4$ & Q$_5$ \\ 
	$\hat{2}_{c}$ &-P$_x$ &-P$_y$ & P$_z$ &-M$_x$ &-M$_y$ & M$_z$ & Q$_1$ & Q$_2$ & Q$_3$ &-Q$_4$ &-Q$_5$ \\
	$\hat{2}_{x}$ & P$_x$ &-P$_y$ &-P$_z$ & M$_x$ &-M$_y$ &-M$_z$ & Q$_1$ & Q$_2$ &-Q$_3$ &-Q$_4$ & Q$_5$ \\ 
	$\hat{2}_{y}$ &-P$_x$ & P$_y$ &-P$_z$ &-M$_x$ & M$_y$ &-M$_z$ & Q$_1$ & Q$_2$ &-Q$_3$ & Q$_4$ &-Q$_5$ \\
	$\hat{4}_{z}^+$ & P$_y$ &-P$_x$ & P$_z$ & M$_y$ &-M$_x$ & M$_z$ &-Q$_1$ & Q$_2$ &-Q$_3$ & Q$_5$ &-Q$_4$ \\ 
	$\hat{4}_{z}^-$ &-P$_y$ & P$_y$ & P$_z$ &-M$_y$ & M$_x$ & M$_z$ &-Q$_1$ & Q$_2$ &-Q$_3$ &-Q$_5$ & Q$_4$ \\
	$\hat{2}_{a}$ & P$_y$ & P$_x$ &-P$_z$ & M$_y$ & M$_x$ &-M$_z$ &-Q$_1$ & Q$_2$ & Q$_3$ &-Q$_5$ &-Q$_4$ \\ 
	$\hat{2}_{b}$ &-P$_y$ &-P$_x$ &-P$_z$ &-M$_y$ &-M$_x$ &-M$_z$ &-Q$_1$ & Q$_2$ & Q$_3$ & Q$_5$ & Q$_4$ \\
	$\overline{1}$  		 &-P$_x$ &-P$_y$ &-P$_z$ & M$_x$ & M$_y$ & M$_z$ & Q$_1$ & Q$_2$ & Q$_3$ & Q$_4$ & Q$_5$ \\
	$\hat{m}_z$		 & P$_x$ & P$_y$ &-P$_z$ &-M$_x$ &-M$_y$ & M$_z$ & Q$_1$ & Q$_2$ & Q$_3$ &-Q$_4$ &-Q$_5$ \\
	$\hat{m}_x$		 &-P$_x$ & P$_y$ & P$_z$ & M$_x$ &-M$_y$ &-M$_z$ & Q$_1$ & Q$_2$ &-Q$_3$ &-Q$_4$ & Q$_5$ \\
	$\hat{m}_y$		 & P$_x$ &-P$_y$ & P$_z$ &-M$_x$ & M$_y$ &-M$_z$ & Q$_1$ & Q$_2$ &-Q$_3$ & Q$_4$ &-Q$_5$ \\
	$\overline{4}_{z}^+$ &-P$_y$ & P$_x$ &-P$_z$ & M$_y$ &-M$_x$ & M$_z$ &-Q$_1$ & Q$_2$ &-Q$_3$ & Q$_5$ &-Q$_4$ \\ 
	$\overline{4}_{z}^-$ & P$_y$ &-P$_y$ &-P$_z$ &-M$_y$ & M$_x$ & M$_z$ &-Q$_1$ & Q$_2$ &-Q$_3$ &-Q$_5$ & Q$_4$ \\
	$\hat{m}_{a}$ &-P$_y$ &-P$_x$ & P$_z$ & M$_y$ & M$_x$ &-M$_z$ &-Q$_1$ & Q$_2$ & Q$_3$ &-Q$_5$ &-Q$_4$ \\ 
	$\hat{m}_{b}$ & P$_y$ & P$_x$ & P$_z$ &-M$_y$ &-M$_x$ &-M$_z$ &-Q$_1$ & Q$_2$ & Q$_3$ & Q$_5$ & Q$_4$ \\
	$1'$             &	P$_x^*$ & P$_y^*$ & P$_z^*$ &-M$_x^*$ &-M$_y^*$ &-M$_z^*$ & Q$_1^*$ & Q$_2^*$ & Q$_3^*$ & Q$_4^*$ & Q$_5^*$ \\	
  $\hat{2}_{z}'$ &-P$_x^*$ &-P$_y^*$ & P$_z^*$ & M$_x^*$ & M$_y^*$ &-M$_z^*$ & Q$_1^*$ & Q$_2^*$ & Q$_3^*$ &-Q$_4^*$ &-Q$_5^*$ \\	 
  $\hat{2}_{x}'$ & P$_x^*$ &-P$_y^*$ &-P$_z^*$ &-M$_x^*$ & M$_y^*$ & M$_z^*$ & Q$_1^*$ & Q$_2^*$ &-Q$_3^*$ &-Q$_4^*$ & Q$_5^*$ \\	 
  $\hat{2}_{y}'$ &-P$_x^*$ & P$_y^*$ &-P$_z^*$ & M$_x^*$ &-M$_y^*$ & M$_z^*$ & Q$_1^*$ & Q$_2^*$ &-Q$_3^*$ & Q$_4^*$ &-Q$_5^*$ \\	
	$\hat{4}_{z}^+1'$ & P$_y^*$ &-P$_x^*$ & P$_z^*$ &-M$_y^*$ & M$_x^*$ &-M$_z^*$ &-Q$_1^*$ & Q$_2^*$ &-Q$_3^*$ & Q$_5^*$ &-Q$_4^*$ \\ 
	$\hat{4}_{z}^-1'$ &-P$_y^*$ & P$_x^*$ & P$_z^*$ & M$_y^*$ &-M$_x^*$ &-M$_z^*$ &-Q$_1^*$ & Q$_2^*$ &-Q$_3^*$ &-Q$_5^*$ & Q$_4^*$ \\
	$\hat{2}_{a}'$ & P$_y^*$ & P$_x^*$ &-P$_z^*$ &-M$_y^*$ &-M$_x^*$ & M$_z^*$ &-Q$_1^*$ & Q$_2^*$ & Q$_3^*$ &-Q$_5^*$ &-Q$_4^*$ \\ 
	$\hat{2}_{b}'$ &-P$_y^*$ &-P$_x^*$ &-P$_z^*$ & M$_y^*$ & M$_x^*$ & M$_z^*$ &-Q$_1^*$ & Q$_2^*$ & Q$_3^*$ & Q$_5^*$ & Q$_4^*$ \\
	$\overline{1}'$      &-P$_x^*$ &-P$_y^*$ &-P$_z^*$ &-M$_x^*$ &-M$_y^*$ &-M$_z^*$ & Q$_1^*$ & Q$_2^*$ & Q$_3^*$ & Q$_4^*$ & Q$_5^*$ \\	
	$\hat{m}_z'$    & P$_x^*$ & P$_y^*$ &-P$_z^*$ & M$_x^*$ & M$_y^*$ &-M$_z^*$ & Q$_1^*$ & Q$_2^*$ & Q$_3^*$ &-Q$_4^*$ &-Q$_5^*$ \\	
	$\hat{m}_x'$    &-P$_x^*$ & P$_y^*$ & P$_z^*$ &-M$_x^*$ & M$_y^*$ & M$_z^*$ & Q$_1^*$ & Q$_2^*$ &-Q$_3^*$ &-Q$_4^*$ & Q$_5^*$ \\	
	$\hat{m}_y'$    & P$_x^*$ &-P$_y^*$ & P$_z^*$ & M$_x^*$ &-M$_y^*$ & M$_z^*$ & Q$_1^*$ & Q$_2^*$ &-Q$_3^*$ & Q$_4^*$ &-Q$_5^*$ \\	
	$\overline{4}_{z}^+1'$ &-P$_y^*$ & P$_x^*$ &-P$_z^*$ &-M$_y^*$ & M$_x^*$ &-M$_z^*$ &-Q$_1^*$ & Q$_2^*$ &-Q$_3^*$ & Q$_5^*$ &-Q$_4^*$ \\ 
	$\overline{4}_{z}^-1'$ & P$_y^*$ &-P$_x^*$ &-P$_z^*$ & M$_y^*$ &-M$_x^*$ &-M$_z^*$ &-Q$_1^*$ & Q$_2^*$ &-Q$_3^*$ &-Q$_5^*$ & Q$_4^*$ \\
	$\hat{m}_{a}'$ &-P$_y^*$ &-P$_x^*$ & P$_z^*$ &-M$_y^*$ &-M$_x^*$ & M$_z^*$ &-Q$_1^*$ & Q$_2^*$ & Q$_3^*$ &-Q$_5^*$ &-Q$_4^*$ \\ 
	$\hat{m}_{b}'$ & P$_y^*$ & P$_x^*$ & P$_z^*$ & M$_y^*$ & M$_x^*$ & M$_z^*$ &-Q$_1^*$ & Q$_2^*$ & Q$_3^*$ & Q$_5^*$ & Q$_4^*$ \\
 			\colrule
		\end{tabular}
	\end{ruledtabular}
	\label{table1}
\end{table}

Table I should be read as follows. Consider a term, apply the transformation rules of the Table line by line, and then sum up. If the total is zero, the term is not present in the magnetic point group $4/mmm1'$.  Consider, for example, the E1-E1-M1 term $P_xP_yM_z$. The first four lines are always characterized by an even number of minus signs and their sum leads to $+4P_xP_yM_z$. The next four lines are instead characterized by an odd number of minus signs and their sum leads to $-4P_xP_yM_z$. So, the overall sum of the first eight lines is zero. It is easy to check that the next eight sum to zero as well, so that the global sum is zero: this term does not contribute to the SHG signal in the $4/mmm1'$ MPG.
For the other groups, it is sufficient to consider the part of Table I that only contains the symmetry elements of the subgroup.  
In this way, we get the results summarized below.

We remind the conclusion of Section III, that the first (absorption) transition is necessarily M1 or E2, the second (absorption) transition is E1 and the 2$\omega$ (emission) transition is E1. However, in the lists reported below, we did not use this experimental constraint (this constraint is instead used in Section IV). The terms considered here and removed in Section IV are noted by $\overline{M}$ or $\overline{Q}$ (instead of $M$ or $Q$), below.  That is, the overline notation denotes outgoing radiation.

\subsection{Azimuthal scan for the high-temperature magnetic point group $4/mmm1'$}

{\it E1-E1-M1 channel -} Only one linear combination is symmetric under all operations of Table I and therefore allowed in this magnetic point group: $\Im[P_zP_xM_y-P_zP_yM_x]$. This linear combination corresponds to six independent terms in the SHG amplitude (corresponding to all permutations of $P_z$, $P_x$ and $M_y$: $\Im[P_zP_xM_y-P_zP_yM_x]$, $\Im[P_zM_yP_x-P_zM_xP_y]$, etc.). 
We remind that imaginary terms in the E1-E1-M1 channel are non-magnetic (i.e., time-reversal even). 

We can evaluate its azimuthal scan through Eqs.~\ref{polax} and \ref{polaxM1}. For example, in the SS channel the amplitude $A_{SS}=0$ because $[P_zP_xM_y]$ is scalarly coupled to $\epsilon_{Sz}\epsilon_{Sx}(\vec{\epsilon}\times\vec{k})_{Sy}$ and $\epsilon_{Sz}=0$ for both `in' and `out' geometries (see Eq.~\ref{polax1}). The same is true for any permutation, because of the term $P_z \rightarrow\epsilon_{Sz}$. Instead, $A_{SP}$ can be nonzero, provided the $z$ term is associated with the outgoing ($P$-polarized) photon. For example, $[P_zP_xM_y]$ is coupled (using Eqs.~\ref{polax} and \ref{polaxM1} again) to $\epsilon_{Pz}\epsilon_{Sx}(\vec{\epsilon}\times\vec{k})_{Sy} = \sin\theta (\sin\phi)(\cos\theta\sin\phi) = \sin\theta\cos\theta\sin^2\phi$ and $[P_zP_yM_x]$ is coupled to $\epsilon_{Pz}\epsilon_{Sy}(\vec{\epsilon}\times\vec{k})_{Sx} = \sin\theta (-\cos\phi)(\cos\theta\cos\phi)=-\sin\theta\cos\theta\cos^2\phi$. As the invariant term is their difference, we find that $\Im[P_zP_xM_y-P_zP_yM_x]$ behaves like $\sin\theta\cos\theta(\sin^2\phi+\cos^2\phi)=\frac{1}{2}\sin(2\theta)$.

All other terms below are evaluated similarly. We have:

\begin{itemize}
\item $A_{SS} = 0$

\item $A_{PS} = 0$

\item $A_{SP} = \frac{1}{2}\sin(2\theta) \Im[P_zP_xM_y-P_zP_yM_x]$

\item $A_{PP} = \frac{1}{2}\sin(2\theta) \big(\Im[P_zP_xM_y-P_zP_yM_x]+ 6~{\rm permutations}\big)$
\end{itemize}

{\it E1-E1-E2 channel -} Only five terms are allowed in this magnetic point group: $\Re[P_zP_xQ_{xz}+P_zP_yQ_{yz}]$; $\Re[(P_x^2+P_y^2)Q_{3z^2-r^2}]$; $\Re[P_z^2Q_{3z^2-r^2}]$; $\Re[(P_xP_y+P_yP_x)Q_{xy}]$; $\Re[(P_x^2-P_y^2)Q_{x^2-y^2}]$. 
If we consider that the ground state has $x^2-y^2$ symmetry, then $\Re[(P_xP_y+P_yP_x)Q_{xy}]=0$ as shown in Appendix C. Using the same calculations as in the E1-E1-M1 channel above, except for using Eq.~\ref{polaxE2} instead of Eq.~\ref{polaxM1}, we get:

\begin{itemize}
\item $A_{SS} = \sin\theta \sin(4\phi)\Re[(P_x^2-P_y^2)Q_{x^2-y^2}] $

\item $A_{PS} = -\sin\theta\cos^2\theta \sin(4\phi)\Re[(P_x^2-P_y^2)Q_{x^2-y^2}] $

\item $A_{SP} = \frac{1}{2}\sin(2\theta)\big(3\Re[{\overline{Q}}_{3z^2-r^2}P_z^2]- \Re[P_zP_xQ_{xz}+P_zP_yQ_{yz})] + \frac{1}{2}(\cos(4\phi) +1) \Re[(P_x^2-P_y^2)Q_{x^2-y^2}] \big)$ 

\item $A_{PP} = \frac{1}{2}(\cos(4\phi) +1) \Re[(P_x^2-P_y^2)Q_{x^2-y^2}] \big) -\frac{1}{2}\sin(2\theta)\cos^2\theta \big(2\Re[{\overline{Q}}_{xz}(P_zP_x+P_xP_z)+{\overline{Q}}_{yz}(P_zP_y+P_yP_z)] - 3\Re[(P_x^2+P_y^2)Q_{3z^2-r^2}] +3 \Re[P_z^2Q_{3z^2-r^2}] +  \frac{3}{2}\sin(2\theta)\Re[P_z^2Q_{3z^2-r^2}] + \sin(2\theta)\sin^2\theta\Re[{\overline{Q}}_{xz}(P_zP_x+P_xP_z)+{\overline{Q}}_{yz}(P_zP_y+P_yP_z)]$ 
\end{itemize}

We remark that the azimuthal ($\phi$) dependence is determined by only one correlation function: $\Re[(P_x^2-P_y^2)Q_{x^2-y^2}]$.

\subsection{Azimuthal scan for the magnetic point group $4/mm'm'$}

The $4/mm'm'$ magnetic group consists of the following symmetry operators: 
$\hat{1}$, $\hat{2}_{z}$, $\hat{2}_{x}'$, $\hat{2}_{y}'$, $\hat{4}_{z}^+$, $\hat{4}_{z}^-$, $\hat{2}_{a}'$, $\hat{2}_{b}'$, $\overline{1}$, $\hat{m}_z$,	$\hat{m}_x'$, $\hat{m}_y'$, $\overline{4}_{z}^+$, $\overline{4}_{z}^-$, $\hat{m}_{a}'$,	$\hat{m}_{b}'$. They are the only ones that have to be considered in Table I to obtain the allowed terms of this MPG.

{\it E1-E1-M1 channel -}  Four terms are now allowed. Apart from $\Im[P_z(P_xM_y-P_yM_x)]$, already present in the HT $4/mmm1'$ MPG, the other three are: $\Re[P_z(P_xM_x+P_yM_y)]$; $\Re[M_z(P_x^2+P_y^2)]$; $\Re[P_z^2M_z]$. As the symmetry reduction is purely magnetic ($4/mm'm'$ and its parent group $4/mmm1'$ have an identical time-reversal even OP), the three real correlation functions are also purely magnetic.
Therefore, we have, from Eqs.~\ref{polax} and \ref{polaxM1}:

\begin{itemize}
\item $A_{SS} = \sin\theta \Re[(P_x^2+P_y^2)M_z]$

\item $A_{PS} = \sin^3\theta \Re[{\overline{M}}_zP_z^2] + \sin\theta \cos^2\theta \Re[{\overline{M}}_z(P_x^2+P_y^2)] -\frac{1}{2} \sin(2\theta) \cos\theta \Re[{\overline{M}}_xP_zP_x+{\overline{M}}_yP_zP_y] - \sin\theta \Re[P_xP_zM_x+P_yP_zM_y)] $

\item $ A_{SP} = 0$

\item $A_{PP} = 0$
\end{itemize}

It is interesting to compare these results with another subgroup of $4/mmm1'$: $4'/mmm'$, which is the one that would be obtained if Varma's orbital current state $\Theta_I$ had been added to the SCOC high-temperature phase.\cite{varma1} 
The main difference between $4/mm'm'$ and $4'/mmm'$ in light of the SHG experiment is due to the fact that two mirror planes of the $4'/mmm'$ point group are not associated with time reversal, thereby failing to describe the experimental data. Though this can be justified just on a symmetry basis, it is interesting to study the technical details that lead to the different azimuthal dependency for the two groups. 
If we consider for example the PP channel, its contribution is given by $\Re[P_z(P_xM_x-P_yM_y)]$ for the $4'/mmm'$ group, and by $\Re[P_z(P_xM_x+P_yM_y)]$ for the $4/mm'm'$ group, with a sum instead of a difference.
The azimuthal dependency of each term are $\Re[P_zP_xM_x]\sim \frac{1}{4}\sin(2\theta)\sin(2\phi)$ and $\Re[P_zP_yM_y] \sim -\frac{1}{4}\sin(2\theta)\sin(2\phi)$. Their linear combination is zero only with the coefficients of the $4/mm'm'$ magnetic group. It is for this reason that $4/mm'm'$ is the only MPG satisfying the experimental constraint 1), 2) and 3) among the highest symmetry magnetic subgroups of $4/mmm1'$.

{\it E1-E1-E2 channel -} In this case eight terms are allowed in the $4/mm'm'$ magnetic point group. Apart from the five common high-temperature terms, that are time-reversal even, the remaining three time-reversal odd terms that are allowed are: $\Im[P_xP_zQ_{yz}-P_yP_zQ_{xz}]$; $\Im[(P_xP_y+P_yP_x)Q_{x^2-y^2}]$; $\Im[(P_x^2-P_y^2)Q_{xy}]$.

As above, we can evaluate their azimuthal scan from Eqs.~\ref{polax} and \ref{polaxE2}:

\begin{itemize}
\item $A_{SS} = \sin\theta \big(\cos^2(2\phi) \Im[(P_x^2-P_y^2)Q_{xy}] - \sin^2(2\phi) \Im[(P_xP_y+P_yP_x)Q_{x^2-y^2}]  \big)$

\item $A_{PS} = -\frac{1}{2}\sin(2\theta)\cos\theta \big(\cos^2(2\phi) \Im[(P_x^2-P_y^2)Q_{xy}] - \sin^2(2\phi) \Im[(P_xP_y+P_yP_x)Q_{x^2-y^2}]  \big) - \sin\theta \cos^2\theta \Im[P_xP_zQ_{yz}-P_yP_zQ_{xz}]$

\item $A_{SP} =-\frac{1}{4}\sin(2\theta) \sin(4\phi) \big(\Im[(P_x^2-P_y^2)Q_{xy}] + \Im[(P_xP_y+P_yP_x)Q_{x^2-y^2}] \big)$ 

\item $A_{PP} = \frac{1}{4}\sin(2\theta)\cos^2\theta \sin(4\phi) \big(\Im[(P_x^2-P_y^2)Q_{xy}] + \Im[(P_xP_y+P_yP_x)Q_{x^2-y^2}] \big)$
\end{itemize}

In this case, it is clear that the conditions 1), 2) and 3) of Section IV are all simultaneously satisfied only if the two tensors $\Im[(P_x^2-P_y^2)Q_{xy}]$ and $\Im[(P_xP_y+P_yP_x)Q_{x^2-y^2}]$ are opposite in value. This will be shown in Appendix C.

\subsection{Azimuthal scan for the magnetic point group $4/m1'$}

The $4/m$ magnetic group consists of the following symmetry operators: 
$\hat{1}$, $\hat{2}_{z}$, $\hat{4}_{z}^+$, $\hat{4}_{z}^-$, $\overline{1}$, $\hat{m}_z$, $\overline{4}_{z}^+$, $\overline{4}_{z}^-$, plus these same eight operators multiplied by the time-reversal symmetry $1'$. As in this case, the symmetry is broken by a non-magnetic, time-reversal even OP, all the allowed terms are time-reversal even, both in the E1-E1-M1 channel (they are therefore imaginary) and in the E1-E1-E2 channel (they are therefore real).

{\it E1-E1-M1 channel -} Only four terms are different from zero. Apart from the high-temperature term common to all subgroups of $4/mmm1'$, which is $\Im[P_zP_xM_y-P_zP_yM_x]$, the three remaining are: $\Im[P_zP_xM_x+P_zP_yM_y]$; $\Im[(P_x^2+P_y^2)M_z]$; $\Im[P_z^2M_z]$. 

{\it E1-E1-E2 channel -} Only eight are different from zero. If we exclude the five high-temperature terms of $4/mmm1'$ symmetry, the three remaining are: $\Re[P_zP_xQ_{yz}-P_zP_yQ_{xz}]$; $\Re[(P_xP_y+P_yP_x)Q_{x^2-y^2}]$; $\Re[(P_x^2-P_y^2)Q_{xy}]$.

It turns out that the terms in the E1-E1-M1 channel correspond to the imaginary (time-reversal even) terms of those associated with $4/mm'm'$ and the terms in the E1-E1-E2 channel correspond to the real (time-reversal even) terms of those associated with $4/mm'm'$. So, apart from the opposite time-reversal behavior, their azimuthal scan will be the same and we shall not report it here. 
We should just change $\Re\leftrightarrow \Im$ in the azimuthal scan of $4/mm'm'$.

Of course, the same comments apply, in particular the E1-E1-M1 channel is compatible with the SHG experimental outcomes 1), 2) and 3), as it was for the $4/mm'm'$ MPG, with the same drawback of having two different correlation functions for SS and PS geometry, with intermediate states at 1.5 eV of both $xy$ and $xz,yz$ symmetries.
Analogously, in the E1-E1-E2 channel the extra condition to match the experimental data should be applied to the real OP: $\Re[(P_x^2-P_y^2)Q_{xy}]=-\Re[(P_xP_y+P_yP_x)Q_{x^2-y^2}]$. However, as demonstrated in Appendix C, this is not the case: rather, $\Re[(P_x^2-P_y^2)Q_{xy}]=+\Re[(P_xP_y+P_yP_x)Q_{x^2-y^2}]$, thereby eliminating this possibility to explain the experimental data.

\section{Calculation of some E1-E1-M1 and E1-E1-E2 transition-matrix elements}

The main advantage of the quantum-mechanical formulation of SHG introduced in Section III, compared to the semiclassical approach in terms of non-linear susceptibilities $\chi$, is that we can calculate the correlation functions and have information about the ground state and intermediate states. 
For example, suppose that the $4/mmm1'$ MPG is characterized by a ground state of $x^2-y^2$ symmetry. We have seen that the first transition stays within the $l=2$ manifold. We can therefore apply the Wigner-Eckhart theorem for the angular part within this subspace (by projecting out possible $l=1,3$ terms): for example, $Q_{x^2-y^2}|x^2-y^2 \rangle \propto |3z^2-r^2 \rangle$.

Suppose we evaluate $\Im[(P_x^2-P_y^2)Q_{xy}]$ when the ground state has $x^2-y^2$ symmetry, i.e., with a $4/mmm1'$ MPG. We remind that the notation $\Im[(P_x^2-P_y^2)Q_{xy}]$ is shorthand for: $\sum_{l,n}\Delta_{l,n}\big(\langle g|P_x |l \rangle \langle l| P_x |n \rangle \langle n| Q_{xy}|g\rangle - \langle g|P_y |l \rangle \langle l| P_y |n \rangle \langle n| Q_{xy}|g\rangle - c.c.\big)$, where $\Delta_{l,n}$ is the resonant denominator, $c.c.$ is the complex conjugate, $|g\rangle$ is the ground state and $|n\rangle$ and $|l\rangle$ are the intermediate states. From the Wigner-Eckhart theorem, $Q_{xy}|x^2-y^2 \rangle  = 0$, as expected, because $|x^2-y^2 \rangle$ has the full symmetry of $4/mmm1'$ and we know from Appendix B that $\Im[(P_x^2-P_y^2)Q_{xy}]\neq 0$ only with the $4/mm'm'$ MPG, not with the $4/mmm1'$ MPG. For the same reason, also $\Re[(P_xP_y+P_yP_x)Q_{xy}]=0$, as already used in Appendix B.1. 

This approach also suggests that, in order to have a non-zero value for these transition-matrix elements, we should have a ground state, for example, of the kind: $|g_{LT}\rangle = (a|x^2-y^2\rangle +i\gamma|xy\rangle)/\sqrt{N}$. Here $N$ is the normalization, and $a$ and $\gamma$ are weights \cite{note2} (with $ \gamma \ll a$). We remark that the $|xy\rangle$ state needs to be imaginary for the $4/mm'm'$ MPG. If it were real, the MPG would have been $4/m1'$. 
In the imaginary case, we can demonstrate the fundamental relation used in Section IV  for the MPG $4/mm'm'$: $\chi_{LT3}=-\chi_{LT4}$, or $\Im[(P_xP_y+P_yP_x)Q_{x^2-y^2}] = - \Im[(P_x^2-P_y^2)Q_{xy}]$.

Evaluate $Q_{xy}|g_{LT}\rangle$. This gives $|n\rangle = 2\sqrt{\frac{3}{N}} i \gamma |3z^2-r^2\rangle$. Then we evaluate $|l \rangle =x|n\rangle= \frac{1}{\sqrt{2}}(Y_{1,-1}-Y_{1,1}) |n\rangle = \sqrt{\frac{6}{5N}}i\gamma |x\rangle$ and, finally, the application of the second $P_x$ gives: $P_x|l \rangle =\sqrt{\frac{3}{5N}}i\gamma \left(|x^2-y^2\rangle- \frac{1}{\sqrt{3}} |3z^2-r^2\rangle\right)$. The calculation of the term in $-P_y^2$ leads to the same coefficient for $|x^2-y^2\rangle$ and to the opposite coefficient for $|3z^2-r^2\rangle$, so that their sum is $2i\sqrt{\frac{3}{5}}\frac{\gamma}{\sqrt{N}}|x^2-y^2\rangle$. Finally, projecting over $\frac{a}{\sqrt{N}}\langle x^2-y^2|$ and taking the imaginary part (that doubles it), we end with $\Im[(P_x^2-P_y^2)Q_{xy}]=4i\sqrt{\frac{3}{5}}\frac{a\gamma}{N}$.
An analogous calculation leads to $\Im[(P_xP_y+P_yP_x)Q_{x^2-y^2}]=-4i\sqrt{\frac{3}{5}}\frac{a\gamma}{N}$ and the reason why the sign is opposite can be understood by noting that this time it is the $|x^2-y^2\rangle$ component of the ground state $|g_{LT}\rangle$ that is selected by the $Q_{x^2-y^2}$ operator and that in the end it will project with the $-i\langle xy|$ part of $\langle g_{LT}|$, whose sign is opposite because it is the complex conjugate. This also highlights the importance of time-reversal breaking. 

Interestingly, the equivalent state of the $4/m1'$ MPG, that we could write as $(\tilde{a} |x^2-y^2\rangle +\tilde{\gamma}|xy\rangle)/\sqrt{N}$, i.e., {\it without} the imaginary unit for the $|xy\rangle$ term, would lead instead to $\chi_{LT3}=+\chi_{LT4}$ and no cancellation of the $\phi$-dependence in the LT phase. For this reason the $4/m1'$ MPG in the E1-E1-E2 channel does not allow to describe the experiment.

As a last calculation for the E1-E1-E2 channel in the $4/mm'm'$ MPG, we show how to evaluate $\chi_{LT5}$, used in Section IV.B.1, and why it is zero if there are no intermediate states of $xz,yz$ symmetry. The steps are the same as above: start from the $|g_{LT}\rangle$ state, apply the $P_xP_zQ_{yz}$ and $P_yP_zQ_{xz}$ operators and finally take the imaginary part of the projection on $\langle g_{LT}|$. The result is again proportional to $a\gamma$, i.e., linear in both coefficients of the $|x^2-y^2\rangle$ and $|xy\rangle$ states. Moreover, $Q_{yz}|x^2-y^2\rangle \propto |yz\rangle$, $Q_{xz}|x^2-y^2\rangle \propto |xz\rangle$, $Q_{yz}|xy\rangle \propto |xz\rangle$ and $Q_{xz}|xy\rangle \propto |yz\rangle$, which shows the necessity for intermediate states of $|xz,yz\rangle$ symmetry at 1.5 eV in this case.

Before finishing, we shall also sketch the calculation of the E1-E1-M1 SS and PS terms of $4/mm'm'$ symmetry. In this case, for the SS geometry, we should apply $L_z|g_{LT}\rangle$ which leads to $|n\rangle = (2ai |xy\rangle +2\gamma|x^2-y^2\rangle)/\sqrt{N}$. As $P_x^2+P_y^2$ does not change the character of either the $|xy\rangle$ or $|x^2-y^2\rangle$ states, the transition-matrix element of $\Re[(P_x^2+P_y^2)M_{z}]$ is non-zero. It is real because the imaginary $|xy\rangle$ state in the intermediate state $|n\rangle$ projects to the imaginary $|xy\rangle$ in $|g_{LT}\rangle$ and analogously for the real $|x^2-y^2\rangle$ part.

Analogously, the term $\Re[{\overline{P}}_xP_zM_{x}+{\overline{P}}_yP_zM_{y}]$ of the PS geometry is non-zero, but it passes through a different intermediate state: not $|xy\rangle$ as for the SS geometry, but $|xz\rangle$ or $|yz\rangle$, depending on whether we consider the transition $M_{x} |x^2-y^2\rangle \rightarrow -i|yz\rangle$ or $M_{x} i|xy\rangle \rightarrow -|xz\rangle$ or $M_{y} |x^2-y^2\rangle \rightarrow -i|xz\rangle$ or, finally, $M_{y} i|xy\rangle \rightarrow |yz\rangle$.

This also shows why for an $x^2-y^2$ ground state ($4/mmm1'$ MPG), all OPs present for the $4/mm'm'$ MPG and not for the $4/mmm1'$ MPG are zero. Indeed, the only non-zero OP compatible with the $x^2-y^2$ ground state of $4/mmm1'$ symmetry is the HT one, found in the previous subsection, $\Re[(P_x^2-P_y^2)Q_{x^2-y^2}]$.


\begin{thebibliography}{99}

\bibitem{fiebig}
M. Fiebig, V. V. Pavlov, and R. V. Pisarev, J. Opt. Soc. Am. B {\bf 22}, 96 (2005).
\bibitem{tor}
D. H. Torchinsky, H. Chu, L. Zhao, N. B. Perkins, Y. Sizyuk, T. Qi, G. Cao, and D. Hsieh, Phys. Rev. Lett. {\bf 114}, 096404 (2015).
\bibitem{zhao1}
L. Zhao, D. H. Torchinsky, H. Chu, V. Ivanov, R. Lifshitz, R. Flint, T. Qi, G. Cao, and D. Hsieh, Nature Phys. {\bf 12}, 32 (2016).
\bibitem{zhao2}
L. Zhao, C. A. Belvin, R. Liang, D. A. Bonn, W. N. Hardy, N. P. Armitage, and D. Hsieh, Nature Phys. {\bf 13}, 250 (2017).
\bibitem{harter}
J. W. Harter, Z. Y. Zhao, J.-Q. Yan, D. G. Mandrus, and D. Hsieh, Science {\bf 356}, 295 (2017).
\bibitem{pershan}
P. S. Pershan, Phys. Rev. {\bf 130}, 919 (1963).
\bibitem{muto}
Y. Tanabe, M. Muto, M. Fiebig, and E. Hanamura, Phys. Rev. B {\bf 58}, 8654 (1998).
\bibitem{sa}
D. Sa, R. Valenti, and C. Gros, Eur. Phys. J. B {\bf 14}, 301 (2000).
\bibitem{dmn}
S. Di Matteo and M. R. Norman, Phys. Rev. B {\bf 94}, 075148 (2016).
\bibitem{dmn2}
S. Di Matteo and M. R. Norman, Phys. Rev. B {\bf 96}, 115156 (2017).
\bibitem{ye}
F. Ye, X. Wang, C. Hoffmann, J. Wang, S. Chi, M. Matsuda, B. C. Chakoumakos, J. A. Fernandez-Baca, and G. Cao, Phys. Rev. B {\bf 92}, 201112(R) (2015).
\bibitem{torre}
A. de la Torre, K. L. Seyler, L. Zhao, S. Di Matteo, M. Scheurer, Y. Li, B. Yu, M. Greven, S. Sachdev, M. R. Norman, and D. Hsieh, arXiv:2008.06516 (in press, Nature Physics).
\bibitem{vaknin}
D. Vaknin, S. K. Sinha, C. Stassis, L. L. Miller, and D. C. Johnston, Phys. Rev. B {\bf 41}, 1926 (1990).
\bibitem{deluca}
G. M. De Luca, G. Ghiringhelli, M. Moretti Sala, S. Di Matteo, M. W. Haverkort, H. Berger, V. Bisogni, J. C. Cezar, N. B. Brookes, and M. Salluzzo, Phys. Rev. B {\bf 82}, 214504 (2010).
\bibitem{helium}
M. Farzaneh, X. F. Liu, M. El-Batanouny, and F. C. Chou, Phys. Rev. B {\bf 72}, 085409 (2005).
\bibitem{boulanger}
M.-E. Boulanger, G. Grissonnanche, S. Badoux, A. Allaire, E. Lefrancois, A. Legros, A. Gourgout, M. Dion, C. H. Wang, X. H. Chen, R. Liang, W. N. Hardy, D. A. Bonn, and L. Taillefer, Nature Commun. {\bf 11}, 5325 (2020).
\bibitem{fauque}
B. Fauque., Y. Sidis, V. Hinkov, S. Pailhes, C. T. Lin, X. Chaud, and P. Bourges, Phys. Rev. Lett. {\bf 96}, 197001 (2006).
\bibitem{he1}
Y. He, J. Moore, and C. M. Varma, Phys. Rev. B {\bf 85}, 155106 (2012).
\bibitem{aji}
V. Aji, Y. He, and C. M. Varma, Phys. Rev. B {\bf 87}, 174518 (2013).
\bibitem{he2}
Y. He, P. A. Lee, and C. M. Varma, Phys. Rev. B {\bf 89}, 035119 (2014).
\bibitem{scheurer}
M. S. Scheurer and S. Sachdev, Phys. Rev. B {\bf 98}, 235126 (2018).
\bibitem{santini1}
P. Santini and G. Amoretti, Phys. Rev. Lett. {\bf 85}, 2188 (2000).
\bibitem{igarashi}
T. Nagao and J.-i. Igarashi, Phys. Rev. B {\bf 72}, 174421 (2005).
\bibitem{note1}
We note that in the case of NpO$_2$, a more recent publication (Ref.~\onlinecite{santini2}) pointed towards a magnetic triakontadipole, rather than a magnetic octupole.
\bibitem{santini2}
P. Santini, S. Carretta, N. Magnani, G. Amoretti, and R. Caciuffo, Phys. Rev. Lett. {\bf 97}, 207203 (2006).
\bibitem{grande}
B. Grande and H. Muller-Buschbaum, Z. Anorg. Allg. Chem. {\bf 417}, 68 (1975).
\bibitem{miller}
L. L. Miller, X. L. Wang, S. X. Wang, C. Stassis, D. C. Johnston, J. Faber, Jr. and C.-K. Loong, Phys. Rev. B {\bf 41}, 1921 (1990).
\bibitem{ITC}
{\it International Tables for Crystallography}, Vol. A, ed. T. Hahn (Kluwer, 1992).
\bibitem{Bilbao}
Bilbao Crystallographic Server, https://www.cryst.ehu.es/
\bibitem{Litvin}
D. B. Litvin, Acta Cryst. A {\bf 64}, 419 (2008).
\bibitem{energy}
To be more specific, for the state (matter+radiation) $|\Phi_g\rangle \equiv |g\rangle|2\hbar\omega\rangle$, the energy is $\Sigma_g = E_g + 2\hbar\omega$, where $E_g$ is the ground state energy of the matter alone and $2\hbar\omega$ the final energy of the radiation. The specific form for the energy of intermediate states depends on the resonant/non-resonant character of the transition. This is detailed in Ref.~\onlinecite{dmn}.
\bibitem{boyd}
R. W. Boyd, {\it Nonlinear Optics} (Academic Press, Burlington, MA, 2008).
\bibitem{schumacher}
A. B. Schumacher, J. S. Dodge, M. A. Carnahan, R. A. Kaindl, D. S. Chemla, and L. L. Miller, Phys. Rev. Lett. {\bf 87}, 127006 (2001).
\bibitem{salamon}
D. Salamon, Ran Liu, M. V. Klein, M. A. Karlow, S. L. Cooper, S-W. Cheong, W. C. Lee and D. M. Ginsberg  Phys. Rev. B {\bf 51}, 6617 (1995).
\bibitem{calc}
In Ref.~\onlinecite{salamon}, they found a power-law dependence for the energy of the $A_{2g}$ state versus the in-plane Cu$-$O distance $d$: $E_{A_{2g}}\simeq 1.7\left(1.91/d\right)^{6.075}$ eV. Using $d\sim 1.98$ \AA~for SCOC, we get $E_{A_{2g}}\sim 1.35$ eV.
\bibitem{kuiper}
P. Kuiper, J.-H. Guo, C. Sathe, L.-C. Duda, J. Nordgren, J. J. M. Pothuizen, F. M. F. de Groot, and G. A. Sawatzky, Phys. Rev. Lett. {\bf 80}, 5204 (1998).
\bibitem{moretti}
M. Moretti Sala, V. Bisogni, C. Aruta, G. Balestrino, H. Berger, N. B. Brookes, G. M. De Luca, D. Di Castro, M. Grioni, M. Guarise, P. G. Medaglia, F. Miletto Granozio, M. Minola, P. Perna, M. Radovic, M. Salluzzo, T. Schmitt, K. J. Zhou, L. Braicovich, and G. Ghiringhelli, New J. Phys. {\bf 13}, 043026 (2011).
\bibitem{ghiringhelli}
G. Ghiringhelli, N. B. Brookes, E. Annese, H. Berger, C. Dallera, M. Grioni, L. Perfetti, A. Tagliaferri, and L. Braicovich, Phys. Rev. Lett. {\bf 92}, 117406 (2004).
\bibitem{varma1}
C. Varma, Phys. Rev. B {\bf 55}, 14554 (1997).
\bibitem{lovenich}
R. Lovenich, A. B. Schumacher, J. S. Dodge, D. S. Chemla, and L. L. Miller, Phys. Rev. B {\bf 63}, 235104 (2001).
\bibitem{Faraday}
J. Xia, P. T. Beyersdorf, M. M. Fejer, and A. Kapitulnik, Appl. Phys. Lett. \textbf{89}, 062508 (2006).
\bibitem{note2}
If the physical origin of $\gamma$ were the spin-orbit coupling $\lambda$, we would have $\gamma=\frac{\lambda}{\Delta E}$, where $\Delta E$ is the energy difference between the $xy$ and $x^2-y^2$ holes. Of course, other physical mechanisms might operate and different states than the $xy$ state might take its place.

\end{thebibliography}
\end{document}